\documentclass[10pt]{article}

\usepackage{geometry}   
\usepackage{float}
\usepackage{booktabs}   
\usepackage{array}  
\usepackage{cite}
\usepackage{comment}
\usepackage{amsmath,amssymb,amsfonts,amsthm}
\usepackage{acronym}
\usepackage{makecell}  
\usepackage{multirow}
\usepackage{cases}
\usepackage{graphicx}
\usepackage{textcomp}
\usepackage{booktabs} 
\usepackage{multirow} 
\usepackage[font=small]{caption}
\usepackage[font=tiny]{subcaption}
\usepackage{makecell}
\usepackage{threeparttable}
\usepackage{color}

\definecolor{blue1}{RGB}{0,60,200}

\usepackage[hidelinks]{hyperref}

\usepackage{setspace}
\usepackage[mathscr]{euscript}

\newcommand{\bfg}[1]{\boldsymbol{#1}}
\newcommand{\bfb}[1]{\boldsymbol{\rm #1}}
\newcommand{\T}{^{\intercal}}
\newcommand{\wdt}{\widetilde}
\newcommand{\ii}{\imath}
\newcommand{\As}{{\bfg{\mathcal{A}}}}
\newcommand{\Etdi}{{\bfb F}}
\newcommand{\Atdi}{\bfb G}
\newcommand{\Ctdi}{{\bfb C}}
\newcommand{\Btdi}{\bfb B}
\newcommand{\Ztdi}{{\bfb Z}}

\newcommand{\ytdi}{\bfb y}
\newcommand{\xys}{\bfb x}
\newcommand{\ft}{{\rm f}}
\newcommand{\sw}{{\rm s}}
\newcommand{\reigvmat}{ \bfg V}
\newcommand{\leigvmat}{ \bfg W}
\usepackage{float}

\acrodef{avr}[AVR]{automatic voltage regulator}
\acrodef{pss}[PSS]{power system stabilizer}
\acrodef{dae}[DAE]{differential algebraic equation}
\acrodef{pf}[PF]{participation factor}
\acrodef{sg}[SG]{synchronous generator}
\acrodef{sssa}[SSSA]{small-signal stability analysis}
\acrodef{tds}[TDS]{time-domain simulation}
\acrodef{tm}[TM]{trapezoidal method}
\acrodef{fem}[FEM]{forward Euler method}
\acrodef{bem}[BEM]{backward Euler method}
\acrodef{tg}[TG]{turbine governor}
\acrodef{der}[DER]{distributed energy resource}
\acrodef{ibr}[IBR]{inverter-based resource}
\acrodef{agc}[AGC]{automatic generation control} 
\acrodef{pll}[PLL]{phase-locked loop} 
\acrodef{lte}[LTE]{local truncation error}
\acrodef{dirk}[DIRK]{diagonally implicit Runge-Kutta}

\renewcommand\appendix{%
  \par
  \setcounter{section}{0}%
  \setcounter{subsection}{0}%
  \setcounter{equation}{0}%
  \renewcommand\thesection{\appendixname~\Alph{section}}%
  \renewcommand\theequation{\Alph{section}.\arabic{equation}}%
  \renewcommand\thefigure{\Alph{section}.\arabic{figure}}%
  \renewcommand\thetable{\Alph{section}.\arabic{table}}%
  \addtocontents{toc}{\string\let\string\numberline\string\tmptocnumberline}%
}

\begin{document}

\begin{center}
{\large\bf Matrix Pencil-Based Analysis of Multirate Simulation Schemes}

\vskip.20in

Liya Huang$^a$, 
Georgios Tzounas$^a$$^{*}$ 
\\[2mm]
{\footnotesize
School of Electrical and Electronic Engineering,
University College Dublin, Ireland
\\[5pt]
$^{*}$Corresponding author.
\\E-mail address: georgios.tzounas@ucd.ie

}
\end{center}

{\small
\noindent
\textbf{Abstract} \   This paper focuses on multirate time-domain simulations of power system models. It proposes a matrix pencil-based approach to evaluate the spurious numerical deformation introduced into power system dynamics by a given multirate integration scheme.  Moreover, it considers the problem of multirate partitioning and discusses a strategy for allocating state and algebraic variables to fast and slow subsystems based on modal \acp{pf}. The suitability and features of the proposed approach are illustrated through numerical simulations that assess the accuracy effects of interfacing, as well as various prediction and solution methods.
}
\\[3pt]

%\vskip.1in

\section{Introduction}
\label{sec:intro}

\subsection{Motivation} 

%The dynamics of modern 
Power system dynamics 
%with high penetration of renewable energy sources, storage systems, and flexible loads, 
are complex and span multiple timescales. %\cite{2023micro, milano2018foundations}. 
This is reflected in the %power 
system model, where various components have different time constants, with some exhibiting faster and others slower responses, resulting in a set of stiff nonlinear \acp{dae}.  Efficient and accurate \acf{tds} of such a model is a challenging task.  This paper focuses on the accuracy and numerical stability analysis of multirate methods, a family of numerical techniques that has been discussed in the literature as a means of enhancing power system \ac{tds} \cite{pekarek2004efficient, li2023iteratively, chen2008variable,crow1994multirate}.

%The addition of renewable energy sources makes the stiffness more significant, as their intermittent and variable behavior introduces faster transients, which makes 

\subsection{Literature Review}

%It consists of partitioning and a numerical integration scheme.
%Both present challenges. 
%Partitioning can be static or dynamic. 
%Dynamic may compromise speed.
%One issue in partitioning is how to separate the algebraic variables in fast and slow.
%Reason is that they dont introduce dynamics of finite speed.
%[10] proposed their partioning based on LTEs, but again, this comes at a computational cost.
%The finite dynamics to which algebraic variables are moslty associated with can be seen through their PFs. These PFs are Defined implicitly, through the PFs of states.

%\cite{kato2009multirate,9809793,peiret2018multibody}.  These aspects have been extensively studied across various domains, including power system dynamics.  Applications specific to power systems include efficient partitioning and simulation approaches 

Multirate methods divide system variables across different timescales, simulating each with a corresponding time step size.  The idea is to improve efficiency by using larger step sizes for slowly varying components, and smaller ones to accurately capture fast-changing dynamics \cite{kato2009multirate,9809793,peiret2018multibody}.  Effective implementation requires addressing two key aspects \cite{li2023iteratively, pekarek2004efficient,8352037}: 
partitioning system variables into different timescales; and selecting a numerical scheme to solve the equations and handle the interfacing between partitions. 
 
Multirate partitioning can be performed once, prior to the simulation (statically)~\cite{pekarek2004efficient, crow1994multirate}; or adaptively updated during the simulation (dynamically) based on the activity of variables and metrics such as \acp{lte} \cite{chen2008variable,crow1995multirate}. Dynamic partitioning promises greater accuracy, but is also more complex to implement, while its need for continuous monitoring and frequent updates of partition boundaries increases computational cost \cite{savcenco2014construction}.  Moreover, an important consideration for both static and dynamic partitioning is how to define the timescales of algebraic variables, as these inherently represent \textit{infinitely fast} or \textit{infinitely slow} dynamics.  In general, there is a lack of systematic strategies to address this issue and existing approaches are heuristic.  For instance, we cite \cite{chen2008variable}
wherein algebraic variables are
partitioned based on the rates at which algebraic equations converge during Newton's iterations.  Moreover, a technique to study the link of fast/slow system dynamics with algebraic variables through the definition of appropriate \textit{mode-in-output} \acp{pf} is presented in
\cite{tzounas2019modal}; however, its application to multirate \ac{dae} partitioning remains unexplored.
Finally, apart from the problem of multirate integration within a single simulation framework, partitioning is also crucial in multirate co-simulation, where different subsystems -- such as electromagnetic transient and electromechanical models -- are coupled, often across different software. In such cases, partitioning is typically performed empirically based on the natural decomposition of power system components \cite{lin2012geco, venkatraman2018dynamic,shu2017multirate}.

%Finally, in many cases, the partitioning of variables is predetermined based on the physical characteristics of different components. For instance, in co-simulation, subsystems are solved according to their distinct time scales and dynamic behaviors, which can range from electromagnetic transients (EMT) to electromechanical dynamics   

%, variables associated with EMT components exhibit the fastest dynamics, followed by those related to power electronics controllers. Stator and network dynamics typically evolve much faster than electrical and mechanical dynamics. 

Once partitioned, the power system model is numerically solved using a multirate integration scheme, where fast dynamics depend on the accurate prediction and interpolation of slow variables to maintain consistent coupling and synchronization \cite{gear1974multirate}.
Interfacing between fast and slow components is a particularly challenging issue: inaccurate interpolation can degrade accuracy, while the use of an explicit predictor may require smaller time steps to avoid numerical instabilities.  On the other hand, implicit predictors have better stability properties but require matrix factorizations. These challenges can ultimately compromise efficiency, thus undermining the initial appeal of using a multirate method. Notably, a systematic approach to assess the accuracy and numerical stability of multirate schemes is currently missing. In this vein, recent works propose a \acf{sssa} framework to evaluate the stability and accuracy of integration methods, see \cite{tzounas2022small, tzounas2022delay,bouterakos2025,tajoli2023mode}, but this has not yet been formulated for or applied to multirate methods.

\subsection{Contribution}

The contribution of the paper is twofold, as follows.
\begin{itemize}
\item A matrix pencil-based approach to evaluate in a unified manner the numerical stability and accuracy of multirate \ac{tds} schemes. 
%\item  The analysis of the impact of different predictors on numerical accuracy in multirate method, along with its validation through time-domain simulations.
\item A discussion on how to partition the algebraic variables of power system \acp{dae} for multirate simulation based on mode-in-output~\acp{pf}.
\end{itemize}

\subsection{Paper Organization}

The remainder of the paper is organized as follows. Section~\ref{sec:multirate} provides preliminaries on power system multirate simulation and discusses how the variables can be partitioned into different timescales using \acp{pf}.  The proposed approach to study the numerical stability and accuracy of multirate methods
is presented in Section~\ref{sec:pencil}.  The case study is discussed in Section~\ref{sec:case}. Conclusions are drawn in Section~\ref{sec:conclusion}.

\section{Power System Multirate Simulation}
\label{sec:multirate}

\subsection{DAE System Model}
\label{sec:model}

The short-term stability of a power system is conventionally studied through a \ac{dae} model, as follows \cite{kundur:94}:
\begin{equation}
\begin{aligned}
\bfg{x}' &= \bfg{f}(\bfg{x},\bfg{y}) \\
\bfg 0_{m,1} &= \bfg{g}(\bfg{x},\bfg{y})
\label{eq:dae}
\end{aligned}
\end{equation}
where $\bfg x = \bfg{x}(t) \in \mathbb{R}^n$ and $\bfg{y}= \bfg{y}(t)\in \mathbb{R}^m$ are the state and algebraic variables of the system, respectively; $\bfg{f}:\mathbb{R}^{n + m} \mapsto \mathbb{R}^n$ and $\bfg{g}:\mathbb{R}^{n + m} \mapsto \mathbb{R}{^m}$; $\bfg{0}_{m,1}$ denotes the zero matrix of dimensions $m \times 1$.  

A multirate simulation consists in numerically solving \eqref{eq:dae} by using different time step sizes for variables evolving in different timescales.  Considering the most common case where the system is partitioned in two timescales, i.e., fast and slow, \eqref{eq:dae} can be rewritten as:
%The number of time step sizes used depends on the number of partitions of the system's dynamics, with the most common being to use two time step sizes.
%A large-scale power system includes multi-timescale components can be divided into several subsystems, which means that state variables and algebraic variables can be partitioned into different timescales.    Hence, 
%For simplicity but without loss of generality, in this paper we consider that the power system model \eqref{eq:dae} is partitioned in two timescales 
%Considering a partitioning of system variables in two sets, namely, fast and slow, 
%
\begin{align}
\label{eq:dae:ff}
\bfg{x}_\ft' &= \bfg{f}_\ft(\bfg{x}_\ft, \bfg{x}_\sw, \bfg{y}_\ft, \bfg{y}_\sw) \\
\label{eq:dae:gf}
\bfg{0}_{m_\ft,1} &= \bfg{g}_\ft(\bfg{x}_\ft, \bfg{x}_\sw, \bfg{y}_\ft, \bfg{y}_\sw) \\
\label{eq:dae:fs}
\bfg{x}_\sw' &= \bfg{f}_\sw(\bfg{x}_\ft, \bfg{x}_\sw, \bfg{y}_\ft, \bfg{y}_\sw) \\
\label{eq:dae:gs}
\bfg{0}_{m_\sw, 1} &= \bfg{g}_\sw (\bfg{x}_\ft, \bfg{x}_\sw, \bfg{y}_\ft, \bfg{y}_\sw)
\end{align}
where $\bfg{x}_\ft \in \mathbb{R}^{n_\ft}$ and $\bfg{y}_\ft \in \mathbb{R}^{m_\ft}$ are the state and algebraic variables of the fast partition; $\bfg{x}_\sw \in \mathbb{R}^{n_\sw}$ and $\bfg{y}_\sw \in \mathbb{R}^{m_\sw}$ are the state and algebraic variables of the slow partition; it is $n_\ft + n_\sw = n$ and $m_\ft + m_\sw = m$.  

\subsection{Multirate Numerical Integration}
\label{sec:multirate:form}

\begin{figure}
\begin{center}
\includegraphics[width=0.43\linewidth]{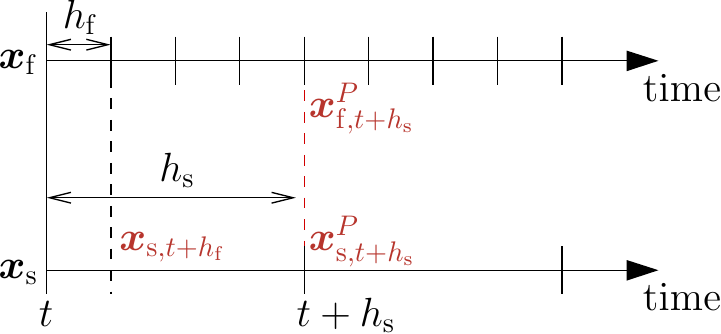}  
\caption{Representation of multirate TDS.}
\label{fig:multirate:scheme}  
\end{center}  
\end{figure}

In this section, we describe a generic multirate integration scheme employed for the solution of \eqref{eq:dae:ff}-\eqref{eq:dae:gs}. Given the model in \eqref{eq:dae:ff}-\eqref{eq:dae:gs}, 
we start from $t=t_0$, and 
fast and slow variables are updated with time steps $h_\ft$ and $h_\sw$, respectively, where $h_\sw > h_\ft$. 
The steps of the multirate scheme, 
a high-level representation of which is given in Fig.~\ref{fig:multirate:scheme}, are as follows.
\begin{enumerate}%[leftmargin=*]
\item Predict the values of $\bfg{x}$ and $\bfg{y}$ at time $t + h_\sw$:
\begin{equation}
    \begin{aligned}
        \bfg{x}_{t + h_\sw}^P &= \bfg{\phi}\left( \bfg{x}_{t},{\bfg{y}_{t},{\bfg{x}_{t + h_\ft}},{\bfg{y}_{t + h_\ft}}, \ldots } \right)\\
        \bfg{0}_{m,1} &= h_\sw \; \bfg{g} \left( \bfg{x}_{t+h_\sw}^P,\bfg{y}_{t+h_\sw}^P \right)  
    \end{aligned}
    \label{eq:alpre}
\end{equation}
where ${^P}$ denotes predicted values; and $\bfg{\phi}\in \mathbb{R}{^{n}}$ is defined by the predictor method adopted.

\item Use the predicted values to interpolate the slow variables at intermediate steps $t + ih_\ft$, $i=1,\dots,r$.  For a generic interpolation method:
%can be expressed as: 
%$\bfg{x}_{t + h_\sw}^P$ and $\bfg{y}_{t + h_\sw}^P$ 
%
\begin{equation}
\begin{aligned}
\bfg{x}_{\sw,t + ih_\ft} &= \bfg{\ii}_\sw ( \bfg{x}_{\sw,t + h_\sw}^P,\bfg{x}_{\sw,t},r ) \\
{\bfg{y}_{\sw,t + ih_\ft}} &= \bfg{\ii}_\sw (\bfg{y}_{\sw,t + h_\sw}^P,\bfg{y}_{\sw,t},r )
\end{aligned}
\end{equation}
where $\bfg{\ii}_\sw\in \mathbb{R}{^{n_\sw}}$; and~$r=h_\sw/h_\ft \in \!\mathbb{N}$.  
\item Integrate the fast equations 
\eqref{eq:dae:ff}-\eqref{eq:dae:gf} with time step $h_\ft$ to calculate the fast variables 
$\bfg{x}_{\ft,t + i h_{\ft}}$, 
$\bfg{y}_{\ft,t + i h_{\ft}}$ at intermediate steps $t+ih_\ft$, $i=1,\dots,r$.

\item Integrate the slow equations \eqref{eq:dae:fs}-\eqref{eq:dae:gs} at $t + h_\sw$ with time step $h_\sw$, using the fast variables $\bfg{x}_{\ft,t + r h_{\ft}}$, 
$\bfg{y}_{\ft,t + r h_{\ft}}$ as inputs.
\item Compare the calculated values of the variables during the integration process with the predicted values.  If $\big | \big | {\bfg{x}_{\sw,t + h_\sw}^P - \bfg{x}_{\sw,t + h_\sw}} \big|\big |_\infty > \! \varepsilon$ or $\big | \big | {\bfg{y}_{\sw,t + h_\sw}^P - \bfg{y}_{\sw,t + h_\sw}} \big|\big |_\infty > \! \varepsilon$, 
then update the predicted values, i.e.,~$\bfg{x}_{\sw,t + h_\sw}^P = {\bfg{x}_{\sw,t + h_\sw}}$ and
$\bfg{y}_{\sw,t + h_\sw}^P = {\bfg{y}_{\sw,t + h_\sw}}$, and return to step 2). 
Otherwise,
%If all variables are within tolerance, 
set $t=t_0+h_\sw$ and return to step 1).
\end{enumerate}

In this paper, we consider that the predictor in 
step~1) may be either explicit or implicit.
For steps 3) and 4) we assume that the integration method employed is implicit. 
As a consequence, the update of variables requires employing an iterative method.
For instance, the $j$-th iteration of Newton’s method employed for the update of fast variables in step~3) is:
\begin{equation}
\bfb{J}_{\ft}
\begin{bmatrix}
\Delta\bfg{x}_{\ft,t + i h_{\ft}}^{(j)} \\
\Delta\bfg{y}_{\ft,t + i h_{\ft}}^{(j)}
\end{bmatrix} =
- 
\begin{bmatrix}
\bfg{F}_{\ft}^{(j)} \\
\bfg{G}_{\ft}^{(j)}
\end{bmatrix}
\label{eq:nt}
\end{equation}
where $\bfb{J}_{\ft} :\mathbb{R}^{(n_\ft + m_\ft)\times(n_\ft + m_\ft)} $ is the Jacobian matrix for the fast subsystem, $\bfg{F}_{\ft}: \mathbb{R}^{n_\ft}$ and $\bfg{G}_{\ft}: \mathbb{R}^{m_\ft}$ are defined by the implicit integration method adopted.  To speed up the solution, a dishonest Newton~(DHN) method can be used, wherein $\bfb{J}_{\ft}$
is factorized only once per time step.  Using DHN in the process above, a summary of the Jacobian factorizations required per time increment $h_\sw$ 
for different integration schemes is given in Table~\ref{tab:fact}. 
%\color{blue1}
We note that a complete assessment of computational efficiency should be based on performance indicators (e.g., wall-clock time). Moreover, meaningful comparisons between different multirate setups require adjusting the time step sizes for each configuration to achieve comparable numerical accuracy. The matrix-pencil-based approach presented in Section~\ref{sec:pencil} can be used to select such step sizes.

%Although computational efficiency of the multirate method cannot be directly inferred from Table~\ref{tab:fact}, these counts primarily reflect the structure of the adopted scheme and are, therefore, independent of system-specific dynamics. A complete efficiency assessment requires other metrics including wall-clock times, linear-solve tallies. 
%However, in order to conduct a fair comparison of different numerical schemes, it is required to use appropriate time step sizes for each scheme to ensure they all achieve the same level of numerical error, which can be achieved by the proposed approach in Section~\ref{sec:pencil}.
\color{black}

\begin{table}[ht]
    \centering
    \renewcommand{\arraystretch}{1.1}
    \caption{Number of Jacobian factorizations every~$h_{\sw}$ using DHN: Single-rate, multirate with explicit predictor, and multirate with implicit predictor.}
    \begin{tabular}{l|cp{2.5mm}cc}
        \toprule\toprule
        & \multicolumn{4}{c}{Matrix order} \\
        Multirate
        & $n \!+\! m$ & 
        $m_\sw$ &
        $n_\ft \!+\! m_\ft$  & $n_\sw \!+\! m_\sw$ \\ 
        \midrule
        No, step $h_{\ft}$ & $r$ &
        0 & 
        0 & 0 \\
        Yes, explicit predictor  & 0 &  
        1 & 
        $r$ &  1 \\
        Yes, implicit predictor
        &  1 & 
        0 & 
        $r $ & 1 \\
        \bottomrule\bottomrule
    \end{tabular}
    \label{tab:fact}
\end{table}

%\color{blue1}
%Computational cost comparison difficult because it depends also on sparsity and fill-in patterns. see KLU algorithm
%\color{black}

\subsection{PF-Based Partitioning} 
\label{sec:pf}

This section focuses on how to define 
\eqref{eq:dae:ff}-\eqref{eq:dae:gs} given \eqref{eq:dae}, i.e.,~on the partitioning of system variables into fast and slow subsets.  As discussed in Section~\ref{sec:intro}, algebraic variables do not define any dynamics of finite speed and are thus more challenging to partition than states.  

To address this issue, \cite{chen2008variable} proposes to evaluate the speed of algebraic variables dynamically, based on the number of Newton iterations required for their equations to converge during the simulation. Variables that converge in a small number of iterations are classified as fast, whereas those requiring more iterations are considered slow. Although this approach may be accurate in some cases, it also has significant limitations.  For example, consider an infinitely slow algebraic variable $y$ defined through the equation $0 = y - y_o$, where $y_o$ is a constant.\footnote{$y$ is infinitely slow as $y' = 0$, or, $T y' = g(y)$ with $T \rightarrow \infty$. Relevant examples in a power system model include, e.g.~auxiliary control variables that define fixed setpoints.}  The equation converges trivially in a single iteration and $y$ is thus misclassified as fast by the above approach, despite being infinitely slow.

%, such as the power setpoint of a \ac{tg} in the absence of secondary frequency control, or the reference voltage of an \ac{avr} in the absence of a stabilizer.}  
%As discussed in \cite{chen2008variable}, the fast-changing algebraic variables will converge quickly when solving the nonlinear algebraic equations by Newton's method. The timescale of the variables are related to the convergence property of the Newton's method, which 
In this paper, we avoid characterizing the speeds of variables -- which is a property of the model -- through purely numerical indicators. 
Instead, we adopt a principled partitioning strategy, based on matrix pencil-based \ac{sssa} and \acp{pf}.
Linearization of \eqref{eq:dae} at an equilibrium $\left( \bfg{x}_{o}, \bfg{y}_{o} \right):= [\bfg x_o\T \ \bfg y_o\T]\T$ gives:
\begin{equation}
\begin{aligned}
 \wdt{\bfg{x}}' & =  \bfg{f}_x \wdt{\bfg{x}} + \bfg{f}_y 
  \wdt{\bfg{y}} \\
\bfg{0}_{m,1} & =  \bfg{g}_x \wdt{\bfg{x}} + \bfg{g}_y \wdt{\bfg{y}}
\label{eq:linear}
\end{aligned}
\end{equation}
where $\wdt{\bfg{x}} = \bfg{x}-\bfg{x}_0$, $\wdt{\bfg{y}} = \bfg{y}-\bfg{y}_0$; and $\bfg{f}_x$, $\bfg{f}_y$, $\bfg{g}_x$, $\bfg{g}_y$ are Jacobian matrices at $\bfg (\bfg x_0,\bfg y_0)$ (where $\T$ is the matrix transpose).  System \eqref{eq:linear} can be rewritten as:
\begin{equation}
\bfb{E} \; \xys'  = \bfb{A} \; \xys
\label{eq:mp}
\end{equation}
where $\xys=(\wdt{\bfg{x}}, \wdt{\bfg{y}})$; $\bfg I_n$ is the $n \times n$ identity matrix;~and: 
\begin{align}
\mathbf{E} &=
\begin{bmatrix}
\bfg I_n & {\bfg 0}_{n,m} \\
{\bfg 0}_{m,n} & \bfg 0_{m}    
\end{bmatrix} \ , \quad 
\bfb{A} =
\begin{bmatrix}
\bfg{f}_x & \bfg{f}_y \\
\bfg{g}_x & \bfg{g}_y 
\end{bmatrix}
\label{eq:singular}
\end{align}
where $\bfg 0_{m}$ is the zero matrix of dimensions $m\times m$.
%
%The family of matrices $s\bfb{E}-\bfb{A}$ parameterized by $s \in \mathbb{C}$, is called the matrix pencil of system \eqref{eq:mp}. 
% The matrix pencil $s\bfb{E}-\bfb{A}$ has $n$ finite eigenvalues and $m$ infinite eigenvalues. If ${\bfg{v}_i}$ and ${\bfg{w} _i}$ are the right and left eigenvectors, respectively, associated with an eigenvalue $s _i$, $i=1,2,\ldots,n+m$,%    
% The GEP is described as follows (for simplicity, the subscript $i$ is omitted):
% \begin{align}
   % \left( s \bfb{E} - \bfb{A} \right)\bfg{v} &= \bfg{0}    \\ 
   % \bfg{w} \left( s \bfb{E} - \bfb{A} \right) &= \bfg{0} 
    %\label{geq}
%\end{align} 
%The solution of the GEP consists in calculating the $n+m$ eigenvalues and eigenvectors of $s\bfb{E} - \bfb{A}$.  
%\color{blue1}
The algebraic variables can be eliminated under the assumption that the Jacobian matrix $\bfg{g}_y$ is invertible. This assumption comes with no loss of generality, as potential singularities can be always resolved by reformulating the DAE system into a dynamically equivalent form with a non-singular $\bfg{g}_y$ \cite{milano2022power}.  The elimination yields: \color{black}
\begin{equation}
\wdt{\bfg{x}}' = \As \; \wdt{\bfg{x}}
\label{eq:reduced}
\end{equation}
where $\As = \bfg{f}_x - \bfg{f}_y \bfg{g}_y^{-1} \bfg{g}_x$. 
The eigenvalues of $\As$ are then found by solving the algebraic problem \cite{milano2020eigenvalue}:
\begin{equation}
\begin{aligned}
\left(s \bfg I_n - \As \right) \bfg{v} &= \bfg{0}_{n,1}  \\ 
  \bfg{w} \left( s {\bfg I}_n - \As \right) &= \bfg{0}_{1,n} 
\label{eq:lep}
\end{aligned}
\end{equation}
 
Every ${s} \in \mathbb{C}$ satisfying \eqref{eq:lep} is an eigenvalue of $\As$, with $\bfg{v}$ and $\bfg{w}$ being the corresponding right and left eigenvectors. 
Moreover, the right and left modal matrices \eqref{eq:reduced} are $\reigvmat = [ \bfg{v}_1 \, \cdots \,  \bfg{v}_n] $ and $\leigvmat = [ \bfg{w}_1\T \, \cdots \, \bfg{w}_{n}\T]\T$. 
%\color{blue1} 
Provided that the eigenvalues 
%$s_i$, $i =1,2,\dots,n$ 
of $\As$ have equal algebraic and geometric multiplicities, the evolution of the $k$-th element of $\widetilde{\bfg x}$ is:
\begin{equation}
        \wdt{ x}_{k}(t)   
    = \sum_{i=1}^{n} e^{s_i t}  \bfg{w}_i \wdt{\bfg x}(0) v_{k,i} 
\end{equation}
where $v_{k,i}$, $w_{i,k}$ are the $k$-th elements of $\bfg v_i$, $\bfg w_i$, respectively. Exciting in the $k$-th differential equation the $k$-th state, e.g., by applying the initial conditions $\wdt x_k(0) =1$, and $\wdt x_h(0) =0$, $h \neq k$, we get:
\begin{equation}
        \wdt{ x}_{k}(t)   
    = \sum_{i=1}^{n}   w_{i,k} v_{k,i} e^{s_i t}= \sum_{i=1}^{n} p_{ki}^{[x]} e^{s_i t} 
    \label{eq:xsolution}
\end{equation}
where $ p_{ki}^{[x]} = w_{i,k} v_{k,i}$ is called mode-in-state \ac{pf} and
provides a measure of the contribution of the $i$-th eigenvalue $s_i$ to the $k$-th state variation $\wdt{x}_k$.
Then, the system's state participation matrix $\bfb{P}_x\in \mathbb{R}^{n \times n}$ is defined as: 
\begin{equation}
  \bfb{P}_x = (p_{ki}^{[x]})_{1\leq (k,i)\leq n}  = \leigvmat\T \circ \reigvmat 
    \label{P_x}
\end{equation}
where $\circ$ denotes the Hadamard product.  
%The $k$-th row, $i$-th column element  \color{blue1} $ p_{ki}^{[x]}$ \color{black} of $\bfb{P}_x$
Given $\bfb{P}_x$, the participation matrix of algebraic variables $\bfb{P}_{y}\in \mathbb{R}^{m \times n}$ is defined as \cite{tzounas2019modal}:
\begin{equation}
\bfb{P}_{y} = -\bfg{g}_y^{-1}\bfg{g}_x \bfb {P}_x 
\label{eq:P_y}
\end{equation}
%
%The rationale of \eqref{eq:P_y} can be found in \cite{tzounas2019modal}.  

The derivation of \eqref{eq:P_y} is provided in %the 
\ref{sec:appA}. \color{black} To ensure that \acp{pf} of different algebraic variables are comparable, each row of $\bfb{P}_{y}$ is normalized by dividing with the Euclidean norm of its entries \cite{moutevelis2024modal}.

%As for algebraic variables, \cite{tzounas2019modal} proposes a method to calculate the participation matrix by constructing a multiple-input multiple-output system whose output vector is set as $\bfg{y}$. The participation matrix of algebraic variables is defined as \cite{milano2020eigenvalue}:
%Similarly, the eigenvalue corresponding to the maximum value in each column of $\bfb P_y$ can be identified.

In this paper, state and algebraic variables are partitioned into fast and slow subsets by using the participation matrices $\bfb{P}_x$ and $\bfb{P}_y$.  
%and used to characterize the timescale of the respective algebraic variable.
%
In particular, by comparing the absolute values of the elements in each row of $\bfb{P}_x$, we identify the eigenvalue that has the largest contribution to each state variable of the system, and denote it as $\lambda_{x,k}$ ($k=1,2,\ldots,n$). Similarly, by comparing the absolute values of the elements in each row of $\bfb{P}_y$, we identify the eigenvalue that has the largest contribution to 
%\color{blue1} 
each algebraic variable,
%\color{black} 
and denote it as $\lambda_{y,j}$ ($j=1,2,\ldots,m$).  Then, based on the natural frequencies of $\lambda_{x,k}$ and $\lambda_{y,j}$, state and algebraic variables are partitioned with a  timescale separation natural angular frequency threshold $\delta$.  For example, an algebraic variable is considered fast if $|\lambda_{y,j}| > \delta$; otherwise, it is slow.

%we calculate the reciprocals of $\|\lambda_{x,i}\|$ and $\| \lambda_{y,k}\|$ as $k_{x,k} = {1}/{\| \lambda _{x,k}\|}$ and $k_{y,j} = {1}/{ \| \lambda _{y,j}\|}$.
%participation matrix quantifies the contribution of each state to each mode of the system. By analyzing the participation matrix, the state mostly influencing an eigenvalue can be identified.
%, allowing to determine which eigenvalue affects a state variable the most. 

%$\reigvmat$ and $\leigvmat$ 
% are right and left eigenvector matrices of the matrix pencil $\hat s\bfg I_n-\mathbf{A}_\sw$. 
%Since the matrix pencil $s\bfg{\mathrm E}-\bfg{\mathrm A}$ has both finite and infinite eigenvalues, assume that the participation matrix $\bfg{\mathrm P}$ for system \eqref{eq:mp} can be formed as follows:
%
%\begin{equation}
%\bfb P = 
%\begin{bmatrix}
%\bfb P_{x,fin} & \bfb P_{x,inf} \\
%\bfb P_{y,fin} & \bfb P_{y,inf} 
%\end{bmatrix}
%\end{equation}
%where the matrices $\bfb P_{x,fin}$, $\bfb P_{x,inf}$, $\bfb P_{y,fin}$, and $\bfb P_{y,inf}$ represent the influence of $\bfg{x}$ and $\bfg{y}$ on finite and infinite eigenvalues, respectively. State variables generally do not contribute to finite eigenvalues, and algebraic variables do not contribute to infinite eigenvalues, making $\bfb P_{x,inf}$ and $\bfb P_{y,fin}$ zero.% Computing eigenvectors for infinite eigenvalues is challenging, but also    unnecessary, as infinite eigenvalues only affect unstable, non-physical components without impacting dynamic behavior.
%\subsubsection{Partitioning Strategy}
%Eigenvalues determine the response characteristics of system variables. 

\section{Numerical Stability and Accuracy} 
\label{sec:pencil}

\subsection{Formulation}
\label{sec:nsssa}

In this section, we present a matrix pencil-based approach to analyze the numerical stability and precision of multirate integration methods.  The main idea is that the spurious numerical deformation introduced to the dynamics of \eqref{eq:dae} by a given multirate scheme can be quantified by studying a linear discrete-time system of the following form:
\begin{equation}
\Etdi_r \; \ytdi_{t+h_\sw} = \Atdi_r \; \ytdi_{t} 
\label{eq:dis}
\end{equation}
where $\Etdi_r$, $\Atdi_r$ and $\ytdi_{t}$ are defined from the specific multirate scheme implemented.  Specifically, \eqref{eq:dis} represents the small-signal dynamics of \eqref{eq:dae} as approximated by the multirate scheme.  The approximated dynamics, in turn, depend on how the system is partitioned, i.e.,~on how \eqref{eq:dae:ff}-\eqref{eq:dae:gs} are defined; as well as on the numerical scheme chosen for the solution of the system, i.e.,~on the specific implementation of steps~1)-5) in Section~\ref{sec:multirate:form}.
The stability of \eqref{eq:dis} can be seen through the properties of the $z$-domain matrix pencil $\widehat z \Etdi_r - \Atdi_r$, where $\widehat z \in \mathbb{C}$.  In particular, \eqref{eq:dis} is asymptotically stable if for every eigenvalue $\widehat z_i$ of $\widehat z \Etdi_r - \Atdi_r$, it holds that $|\widehat z_i| < 1$.  An instability of \eqref{eq:dis} indicates numerical instability of the multirate scheme.
Moreover, the numerical deformation introduced to the dynamic modes of the power system model by the multirate method can be seen by comparing the eigenvalues of $\widehat z \Etdi_r - \Atdi_r$ to those of $s \bfg I_n - \As$.  For a given eigenvalue $s_i$ of $s \bfg I_n - \As$, the relative deformation is $|\widehat s_i - s_i|/|s_i|$
where $\widehat s_i = {\rm log}(\widehat z_i )/h_\ft$ is the corresponding numerically deformed eigenvalue, mapped to the $s$-domain to facilitate comparison.

\subsection{Application to Illustrative Multirate Scheme} 
\label{sec:multirate:illu}

In this section, we illustrate the proposed matrix pencil-based numerical stability and accuracy analysis.  To this end, we consider an implementation of steps~1)-5) in Subsection~\ref{sec:multirate:form} that presents similarities with the scheme adopted in \cite{chen2008variable}.
In particular, we consider prediction in step~1) by the \ac{fem}, and solution of fast and slow variables in steps~3) and 4 by the \ac{tm}.  The process at the time interval $[t_0, t+h_\sw]$ is described as follows. 

In step 1), the values of all variables are predicted using \ac{fem}:
\begin{equation}
\begin{aligned}
\bfg{x}^P_{t+h_\sw} &= \bfg{x}_{t} + h_\sw \bfg{f}(\bfg{x}_{t}, \bfg{y}_{t}) 
\\
\bfg{0}_{m,1} &= h_\sw \bfg{g}(\bfg{x}^P_{t + h_\sw}, \bfg{y}^P_{t + h_\sw})
\end{aligned}
\label{eq:pre}
\end{equation}

In step~2), the values of the slow variables at intermediate steps $t + ih_\ft$ are interpolated linearly:
\begin{equation}
\begin{aligned}
\bfg{x}_{\sw,t + ih_\ft} &= i ( \bfg{x}^P_{\sw,t+h_\sw} - \bfg{x}_{\sw,t} )/r + \bfg{x}_{\sw,t}  \\
\bfg{y}_{\sw,t + ih_\ft} &= i ( \bfg{y}^P_{\sw,t+h_\sw}- \bfg{y}_{\sw,t} )/r + \bfg{y}_{\sw,t}
\end{aligned}
\label{eq:pxy}
\end{equation}
 
In step 3), fast equations at 
$t \!+\! i h_\ft$ are solved with \ac{tm}: 
%
%\begin{equation}
\begin{align}
\nonumber
\bfg{x}_{\ft, t + ih_\ft} 
=& 
\bfg{x}_{\ft,t + (i - 1) h_\ft}  
+ \frac{h_\ft}{2}{\bfg{f}_\ft} 
( \bfg{x}_{t + ih_\ft} ,\bfg{y}_{t + ih_\ft} )
\\
\label{eq:gd12}
& + \frac{h_\ft}{2}{\bfg{f}_\ft}
( {\bfg{x}_{ t + (i - 1) h_\ft},{\bfg{y}_{t + (i - 1)h_\ft }}} 
) 
\\
\nonumber
\bfg{0}_{m_\ft,1} 
=& 
h_\ft  \bfg{g}_\ft ( {{\bfg{x}_{t + ih_\ft}},{\bfg{y}_{t + ih_\ft}}} )
\end{align}
 
In step 4), slow equations at $t \!+\!  h_\sw$ are solved with \ac{tm}: 
\begin{equation}
\begin{aligned}
\bfg{x}_{\sw, t + h_\sw} 
&= 
\bfg{x}_{\sw,t}  
+ 
\frac{h_\sw}{2} \bfg{f}_\sw
( \bfg{x}_{t} ,\bfg{y}_{t} )
+ \frac{h_\sw}{2}{\bfg{f}_\sw}
( {\bfg{x}_{ t +  h_\sw},{\bfg{y}_{t +  h_\sw }}} 
)  
\\
\bfg{0}_{m_\sw,1} 
&= 
h_\sw  \bfg{g}_\sw ( {{\bfg{x}_{t + h_\sw}},{\bfg{y}_{t + h_\sw}}} )
\end{aligned}
\label{eq:gd34}
\end{equation}

We proceed to apply the proposed analysis as described in Section~\ref{sec:nsssa}. To this end, we linearize \eqref{eq:pre}-\eqref{eq:gd34} around the equilibrium $(\bfg x_o, \bfg y_o)$, which is also a fixed point of \eqref{eq:pre}-\eqref{eq:gd34} under the assumption that $
(\bfg{x}_{t+ \tau}, \bfg{y}_{t+ \tau})
= (\bfg x_o, \bfg y_o)$ for $\tau \in [-h_\sw , 0]$.
%\color{blue1}

Linearization of \eqref{eq:pre} gives:
\begin{equation}
\begin{aligned}
\wdt{\bfg{x}}^P_{t+h_\sw} &= \wdt{\bfg{x}}_{t} + {h_\sw}\left( \bfg{f}_x {\wdt{\bfg{x}}_t} + \bfg{f}_y
\wdt{\bfg{y}}_t \right)\\
\bfg 0_{m,1} &= \bfg{g}_x  \wdt{\bfg{x}}^P_{t+h_\sw} +  \bfg{g}_y \wdt{\bfg{y}}^P_{t+h_\sw}
\end{aligned}
\label{eq:li}
\end{equation}
 
Linearization of \eqref{eq:pxy} gives:
\begin{equation}
\begin{aligned}
\wdt{\bfg x}_{\sw,{t} + h_\ft} & = ( \wdt{\bfg x}^P_{\sw,t + h_\sw} - \wdt{\bfg x}_{\sw,t} )/r + \wdt{\bfg x}_{\sw,t} \\
\wdt{\bfg y}_{\sw,t + h_\ft} & = ( \wdt{\bfg y}^P_{\sw,{t} + h_\sw} - \wdt{\bfg y}_{\sw,t} )/r + \wdt{\bfg y}_{\sw,t}
\end{aligned}
\label{eq:li_2}
\end{equation}
 
Linearization of \eqref{eq:gd12} gives:
\begin{align}
\nonumber
\wdt{\bfg x}_{\ft, t + i h_\ft} =& 
\wdt{\bfg x}_{\ft,{t} + (i - 1)h_\ft} 
  + \frac{{h_\ft}}{2}
  ( {{\bfg f_{\ft,x}}
  \wdt{\bfg x}_{t + i h_\ft} + {\bfg f_{\ft,y}} \wdt{\bfg y}_{{t} + ih_\ft}} )
  \\
  \label{li_3}
  & 
 \quad  \!\!\!\!\!
  + \frac{h_\ft}{2} 
  ( {\bfg f_{\ft,x} \wdt{\bfg x}_{{t} + (i - 1)h_\ft} \!\! + \! {\bfg f_{\ft,y}} \wdt{\bfg y}_{{t} + (i - 1)h_\ft}} ) 
  \\
  \bfg 0_{m_\ft,1} & 
= h_\ft
\left( \bfg g_{\ft,x}
\wdt{\bfg x}_{t + i h_\ft} + \bfg g_{\ft,y} \wdt{\bfg y}_{t + i h_\ft} 
\right)
\nonumber
\end{align}
% \begin{align}
% \nonumber
% \wdt{\bfg x}_{\ft, t + i h_\ft} &=  
% \wdt{\bfg x}_{\ft,{t} + (i - 1)h_\ft} 
%   + \frac{{h_\ft}}{2}
%   ( {{\bfg f_{x,\ft}}
%   \wdt{\bfg x}_{t + i h_\ft} + {\bfg f_{y,\ft}} \wdt{\bfg y}_{{t} + ih_\ft}} )
%   + \frac{h_\ft}{2} 
%   ( {\bfg f_{x,\ft} \wdt{\bfg x}_{{t} + (i - 1)h_\ft} \!\! + \! {\bfg f_{y,\ft}} \wdt{\bfg y}_{{t} + (i - 1)h_\ft}} ) 
%   \\
%   \bfg 0_{m_\ft,1} & 
% = h_\ft
% \left( \bfg g_{x,\ft}
% \wdt{\bfg x}_{t + i h_\ft} + \bfg g_{y,\ft} \wdt{\bfg y}_{t + i h_\ft} 
% \right) \label{li_3}
% \end{align}

Finally, linearization of \eqref{eq:gd34} gives:
\begin{align}
\nonumber
\wdt{\bfg x}_{\sw, t + h_\sw} &=
\wdt{\bfg x}_{\sw,t} + \frac{h_\sw}{2}
( \bfg f_{\sw,x} \wdt{\bfg x}_{t + h_\sw} + \bfg f_{\sw,y} \wdt{\bfg y}_{t + h_\sw} )
\\
&  
\hspace{5mm}
+\frac{h_\sw}{2} 
  ( \bfg f_{\sw,x} \wdt{\bfg x}_{t}  +  \bfg f_{\sw,y} \wdt{\bfg y}_{t} ) 
  \label{eq:li_4}
  \\
\bfg 0_{m_\sw,1} & 
= h_\sw
\left( \bfg g_{\sw,x}
\wdt{\bfg x}_{t + h_\sw} + \bfg g_{\sw,y} \wdt{\bfg y}_{t + h_\sw} 
\right)
\nonumber
\end{align}
\color{black}
We provide the following proposition.

\textit{Proposition~1}: The spectral properties of \eqref{eq:li}-\eqref{eq:li_4} can be seen by studying an equivalent discrete-time system in the form of \eqref{eq:dis}, where $\ytdi_{t+h_\sw} = (\xys_{{t} + rh_\ft}, \ldots \xys_{{t} + h_\ft})$, $\ytdi_{t} = (\xys_{t + (r - 1)h_\ft}, \ldots, \xys_{t})$, and $\Atdi_r$, $\Etdi_r$ are properly defined matrices.
 
%\color{blue1}
The proof of Proposition~1 is provided in %the 
\ref{sec:appB} 
%\color{black}
and is written in a general way to accommodate a broader class of predictor and corrector methods through a parameterized representation.
Then, following from this proof, the case study discussed in the next section compares the accuracy of multirate schemes implemented using different predictor and fast/slow variables solution methods.

\begin{comment}
%For instance, if in step~1 we consider prediction by \ac{tm}, then \eqref{eq:pre} changes into:
%\color{red}
%For the implicit predictor in step 1), the values of all variables are predicted using an implicit method, such as the \ac{tm}:
%
\begin{equation}
\begin{aligned}
\bfg{x}^P_{t+h_\sw} =& \bfg{x}_{t} \!+ \!\frac{h_\sw}{2} \bfg{f}(\bfg{x}_{t}, \bfg{y}_{t}) 
\!+\! \frac{h_\sw}{2} \bfg{f}(\bfg{x}_{t+h_\sw}, \bfg{y}_{t+h_\sw}) \\
\bfg{0}_{m,1} =& h_\sw \bfg{g}(\bfg{x}^P_{t + h_\sw}, \bfg{y}^P_{t + h_\sw})
\end{aligned}
\label{eq:backpre}
\end{equation}
%
and linearization gives:
%
\color{red}
\begin{equation}
\begin{aligned}
\wdt{\bfg{x}}^P_{t+h_\sw} &= \wdt{\bfg{x}}_{t} + \frac{h_\sw}{2}\left( \bfg{f}_x {\wdt{\bfg{x}}_{t}} + \bfg{f}_y
\wdt{\bfg{y}}_{t} \right)\\
& \quad + \frac{h_\sw}{2}\left( \bfg{f}_x {\wdt{\bfg{x}}_{t+h_\sw}} + \bfg{f}_y
\wdt{\bfg{y}}_{t+h_\sw} \right)\\
\bfg 0_{m,1} &= \bfg{g}_x  \wdt{\bfg{x}}^P_{t+h_\sw} +  \bfg{g}_y \wdt{\bfg{y}}^P_{t+h_\sw}
\end{aligned}
\label{eq:backli}
\end{equation}
\color{black}
\end{comment}

%The proposed \ac{sssa}-based approach is valid near stationary solutions but remains robust, offering reliable deformation estimates under varying conditions. For similar insights, see, e.g.,~\cite{tzounas2022small}. 

\section{Case Study}
\label{sec:case}

% \begin{figure}
% \begin{center}
% \includegraphics[width=0.5\linewidth]{figs/wscc_eigenvalues.pdf} 
% \caption{\textcolor{blue1}{Eigenvalues of WSCC system: Different values of $\delta $ lead to different system partitioning. }}
% \label{fig:wcss_eig}  
% \end{center}  
% \end{figure}

% \begin{figure}[ht!]
%     \centering
%     \begin{subfigure}[b]{0.55\linewidth}
%         \includegraphics[width=\linewidth]{figs/error_real.pdf}
%         \caption{}
%         \label{fig:sub1}
%     \end{subfigure}
    
%     %\vspace{0.1cm} % 调整两个子图之间的垂直间距
    
%     \begin{subfigure}[b]{0.55\linewidth}
%         \includegraphics[width=\linewidth]{figs/error_im.pdf}
%         \caption{}
%         \label{fig:sub2}
%     \end{subfigure}
    
%     \caption{\textcolor{blue1}{}}
%     \label{fig:error_reim}
% \end{figure}

This section presents simulation results based on the well-known WSCC 9-bus benchmark system \cite{sauer2017power}.  The system consists of 6 transmission lines and 3 medium voltage/high voltage transformers; 3 \acp{sg} equipped with \acp{avr}, \acp{pss}, and \acp{tg}.  In total, the system’s \ac{dae} model includes 33 state and 64 algebraic variables.  Simulation results are obtained with Dome \cite{milano2013python}.  Eigenvalues are computed with LAPACK \cite{anderson1990lapack}.

 \begin{table}[ht!]
 \centering
 \caption{PF-based partitioning: relative numerical deformation of dominant mode.}
 \begin{tabular}{clcc}
     \toprule\toprule
     $h_\ft$~[s] & $h_\sw$~[s] & \makecell{PF-based for \\ $\bfg x$ and $\bfg y$} & \makecell{PF-based for $\bfg x$;  \\ ($\bfg y = \bfg y_\ft$)} 
     \\ 
     \midrule
\multirow{3}{*}{$0.001$} & $0.005$ & $0.42$\% & $0.42$\% \\
& $0.01$  & $0.42$\% & $0.42$\% \\
& $0.05$  & $0.42$\% & $0.42$\% \\
\midrule
$0.002$ & \multirow{3}{*}{$0.1$} & $0.84$\% & $0.838$\% \\
$0.004$ &    & $1.68$\% & $1.68$\% \\
$0.005$ &    & $2.10$\% & $2.09$\% \\
\bottomrule\bottomrule
\end{tabular}
\label{Perror}
\end{table}
 
\begin{figure*}[ht!]
    \centering
    \begin{subfigure}{0.325\textwidth}
    \centering
\includegraphics[width=\linewidth]{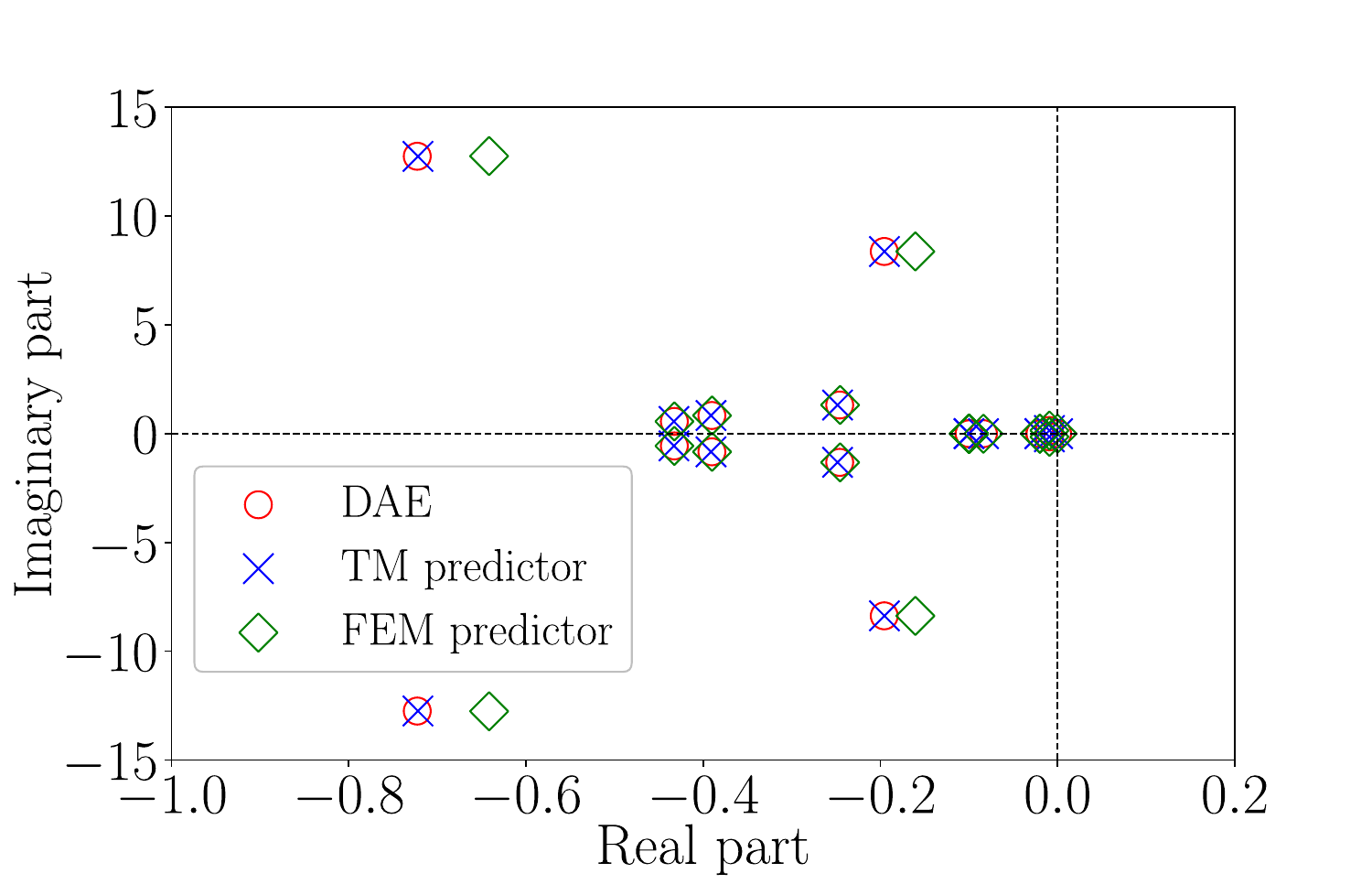}
        \caption{$h_\sw = 0.005$~s, $h_\ft = 0.001$~s.}
        \label{fig:subfig1}
    \end{subfigure}
    \hfill
    \begin{subfigure}{0.325\textwidth}
      \centering  \includegraphics[width=\linewidth]{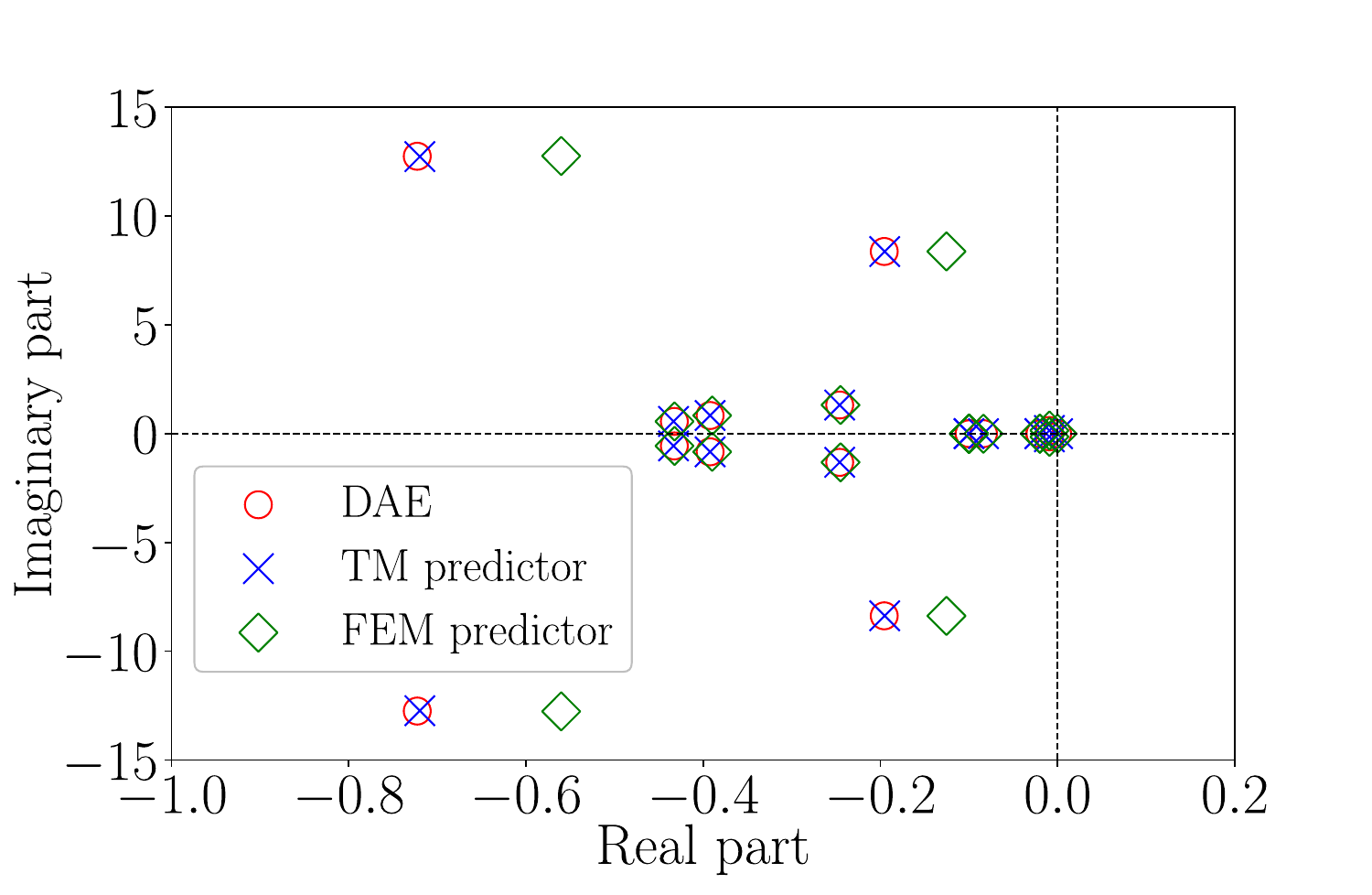}
        \caption{$h_\sw = 0.01$~s, $h_\ft = 0.002$~s.}
        \label{fig:subfig2}
    \end{subfigure}
    \hfill
    \begin{subfigure}{0.325\textwidth}
        \centering    \includegraphics[width=\linewidth]{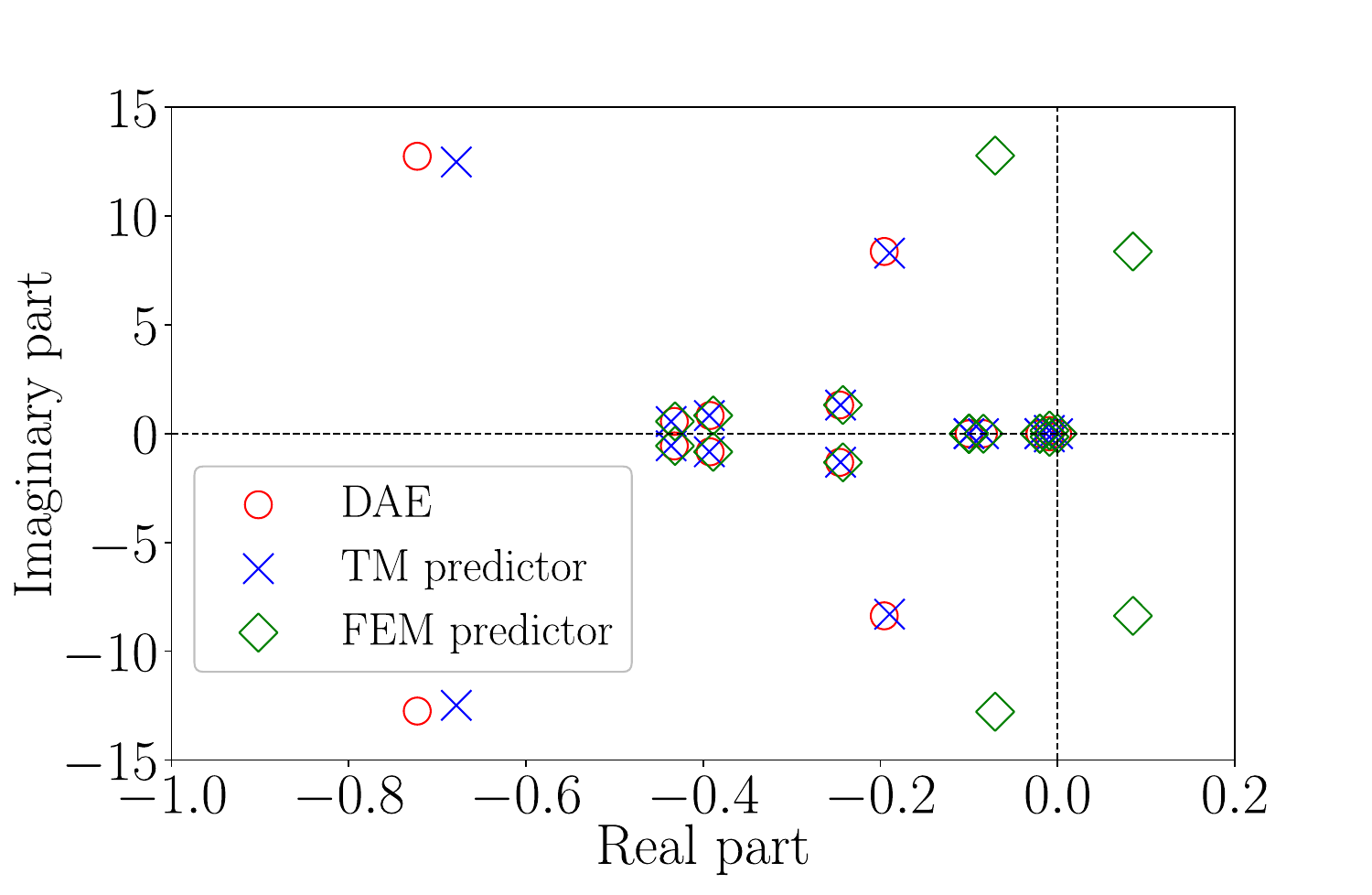}
        \caption{$h_\sw = 0.04$~s, $h_\ft = 0.008$~s.}
        \label{fig:subfig3}
    \end{subfigure}
    \caption{Eigenvalue analysis of 
    multirate schemes:
    \ac{fem} prediction vs. \ac{tm} prediction.}
    \label{fig:predictor}
\end{figure*}

\begin{figure*}[t]
    \centering
    \begin{subfigure}{0.325\textwidth}
        \centering
      \includegraphics[width=\linewidth]{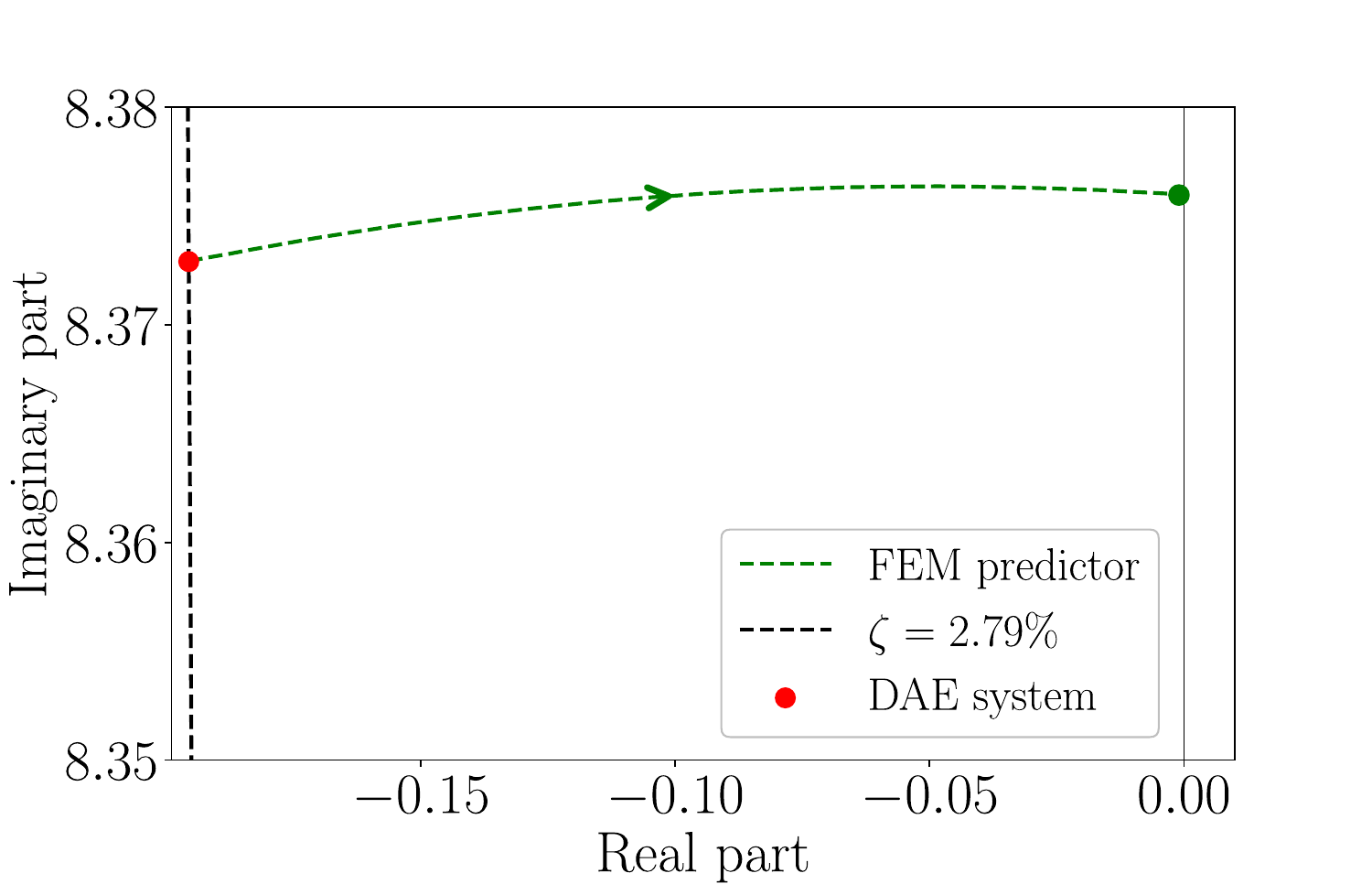}
        \caption{FEM prediction, TM solution.}
        \label{fig:a}
    \end{subfigure}
    \hfill
    \begin{subfigure}{0.325\textwidth}
        \centering
    \includegraphics[width=\linewidth]{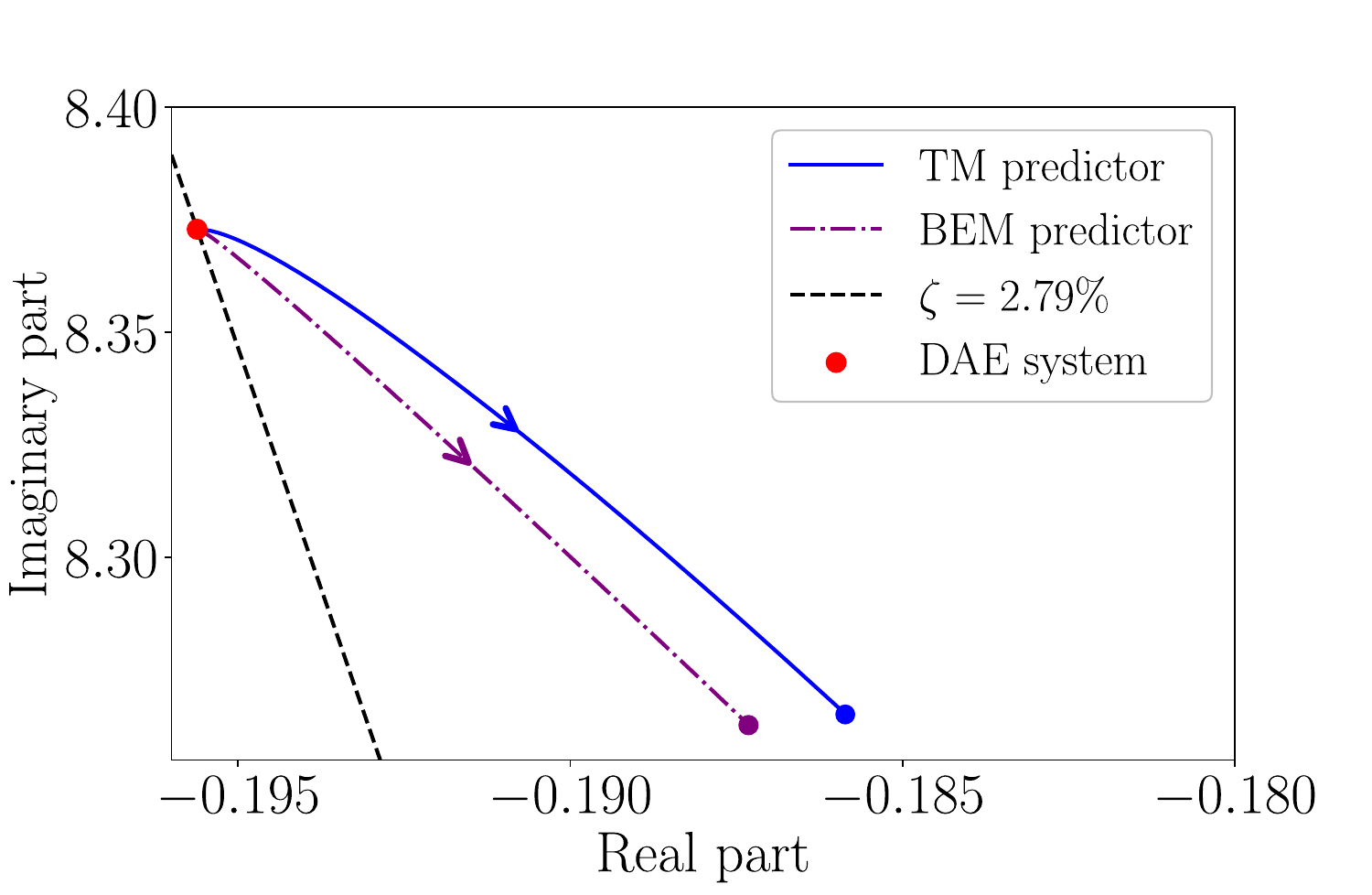}
        \caption{Implicit prediction, TM solution.}
        \label{fig:b}
    \end{subfigure}
    \hfill
    \begin{subfigure}{0.325\textwidth}
        \centering
    \includegraphics[width=\linewidth]{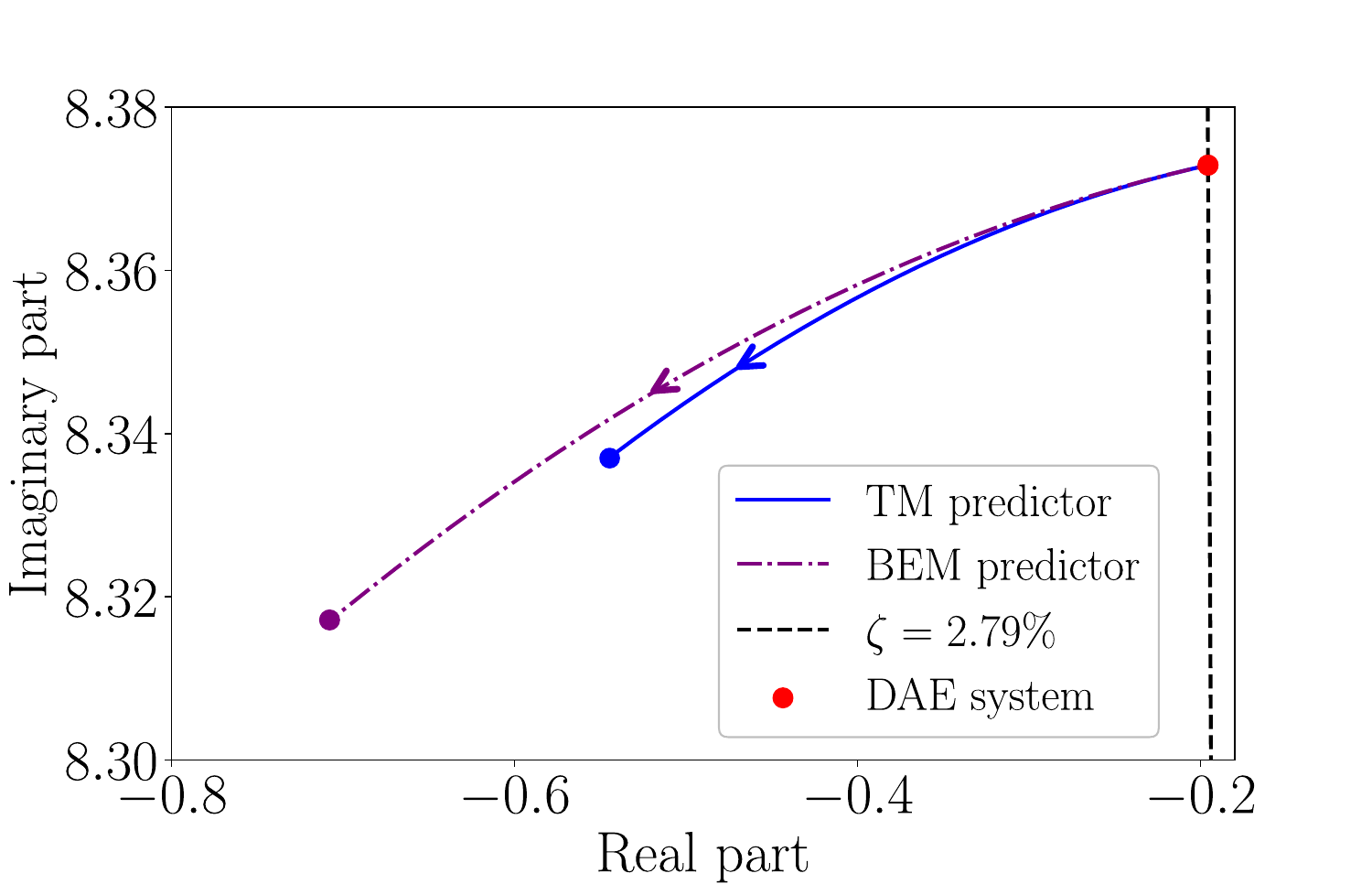}
\caption{Implicit~prediction, BEM solution.}
        \label{fig:c}
    \end{subfigure}
    \caption{Multirate integration: Dominant mode deformation as $h_\ft$ increases from  $10^{-4}$ to 0.005~s, $r = 10$. The points at the end of each line correspond to $h_\ft = 0.005$~s.}
    \label{fig:all_plots}
\end{figure*}

\begin{figure*}[t]
    \centering
    \begin{subfigure}{0.325\textwidth}
        \centering
      \includegraphics[width=\linewidth]{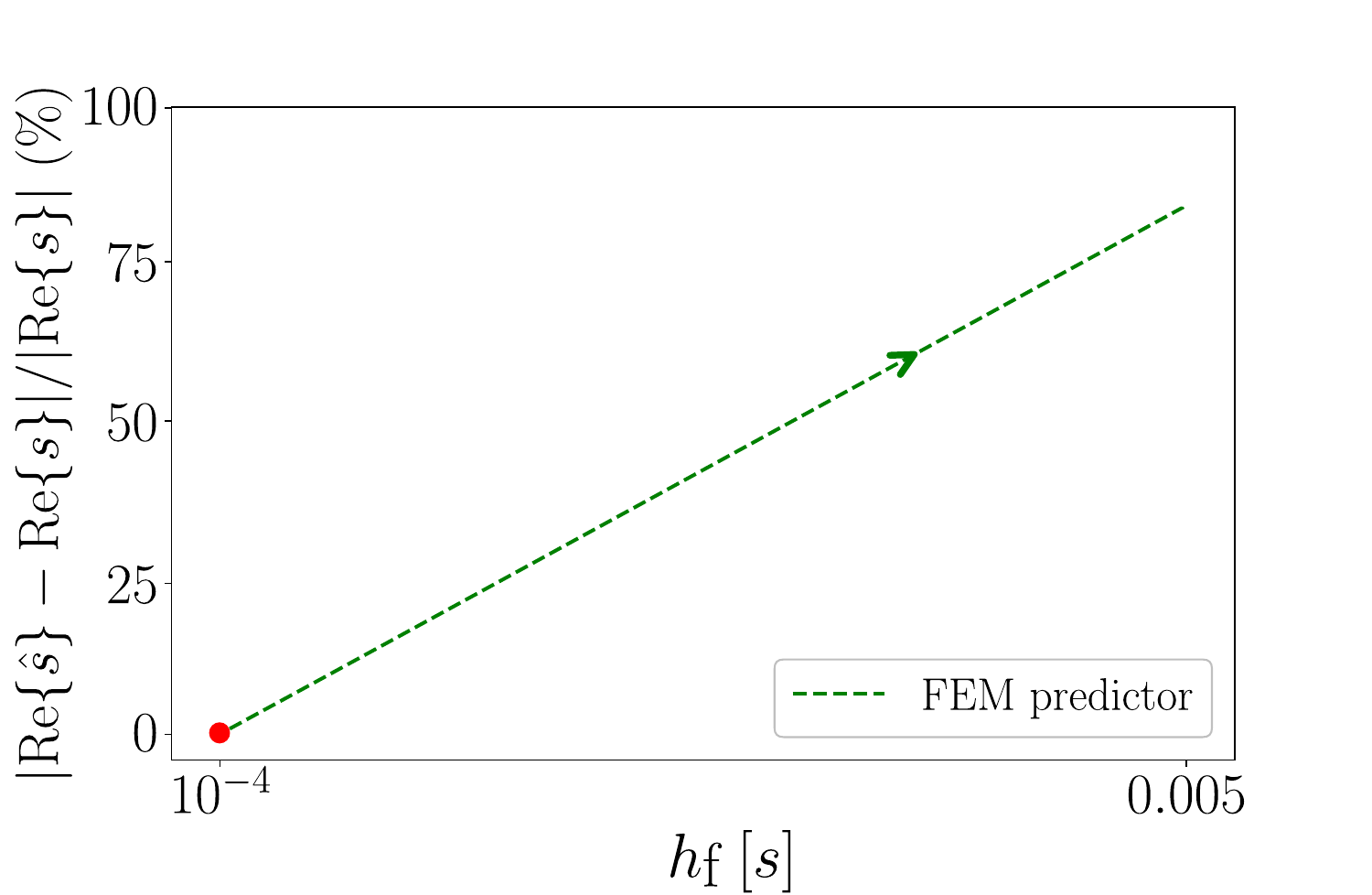}
        \caption{FEM prediction, TM solution.}
        \label{fig:a_re}
    \end{subfigure}
    \hfill
    \begin{subfigure}{0.325\textwidth}
        \centering
    \includegraphics[width=\linewidth]{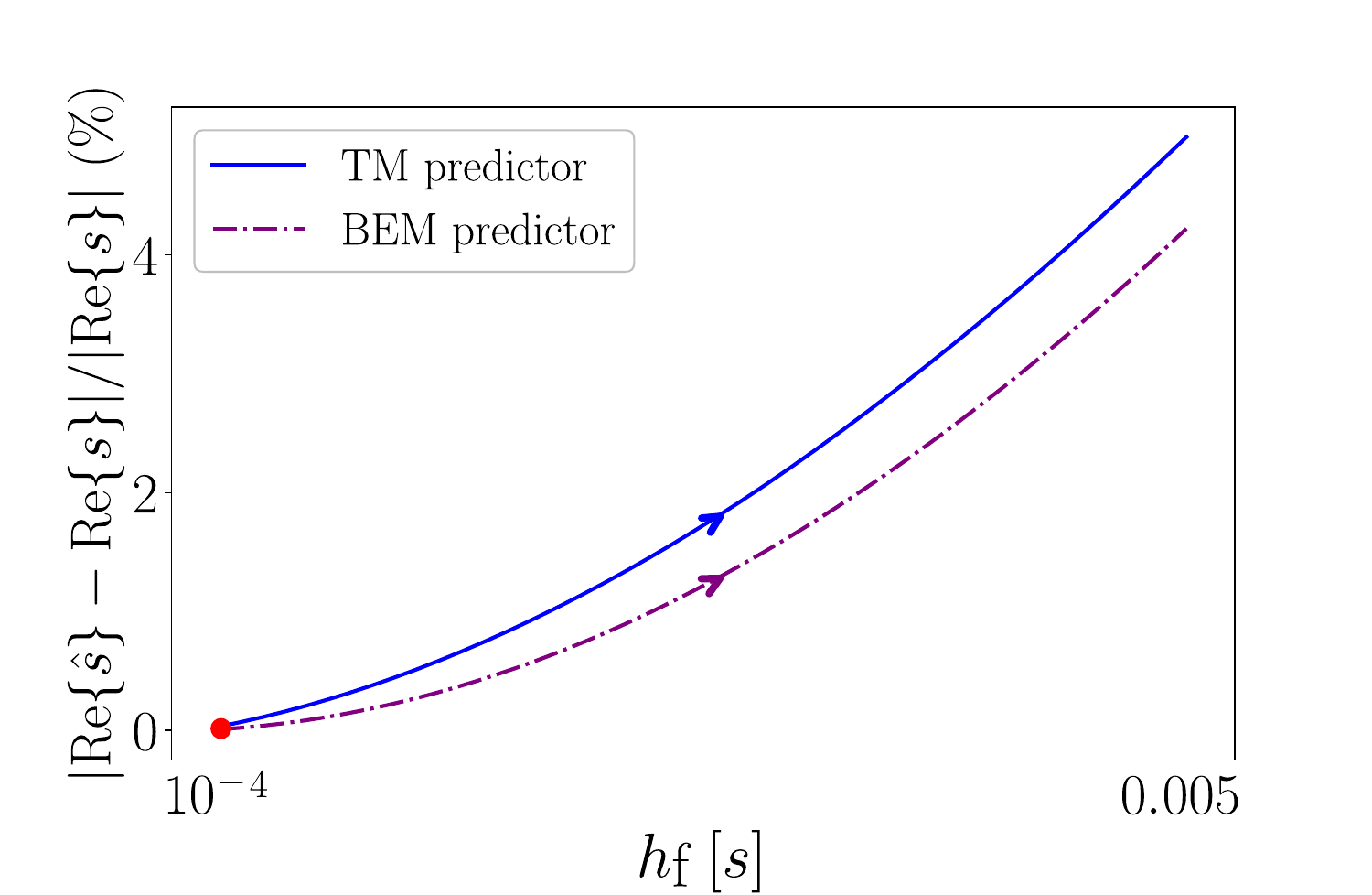}
        \caption{Implicit prediction, TM solution.}
        \label{fig:b_re}
    \end{subfigure}
    \hfill
    \begin{subfigure}{0.325\textwidth}
    \centering
    \includegraphics[width=\linewidth]{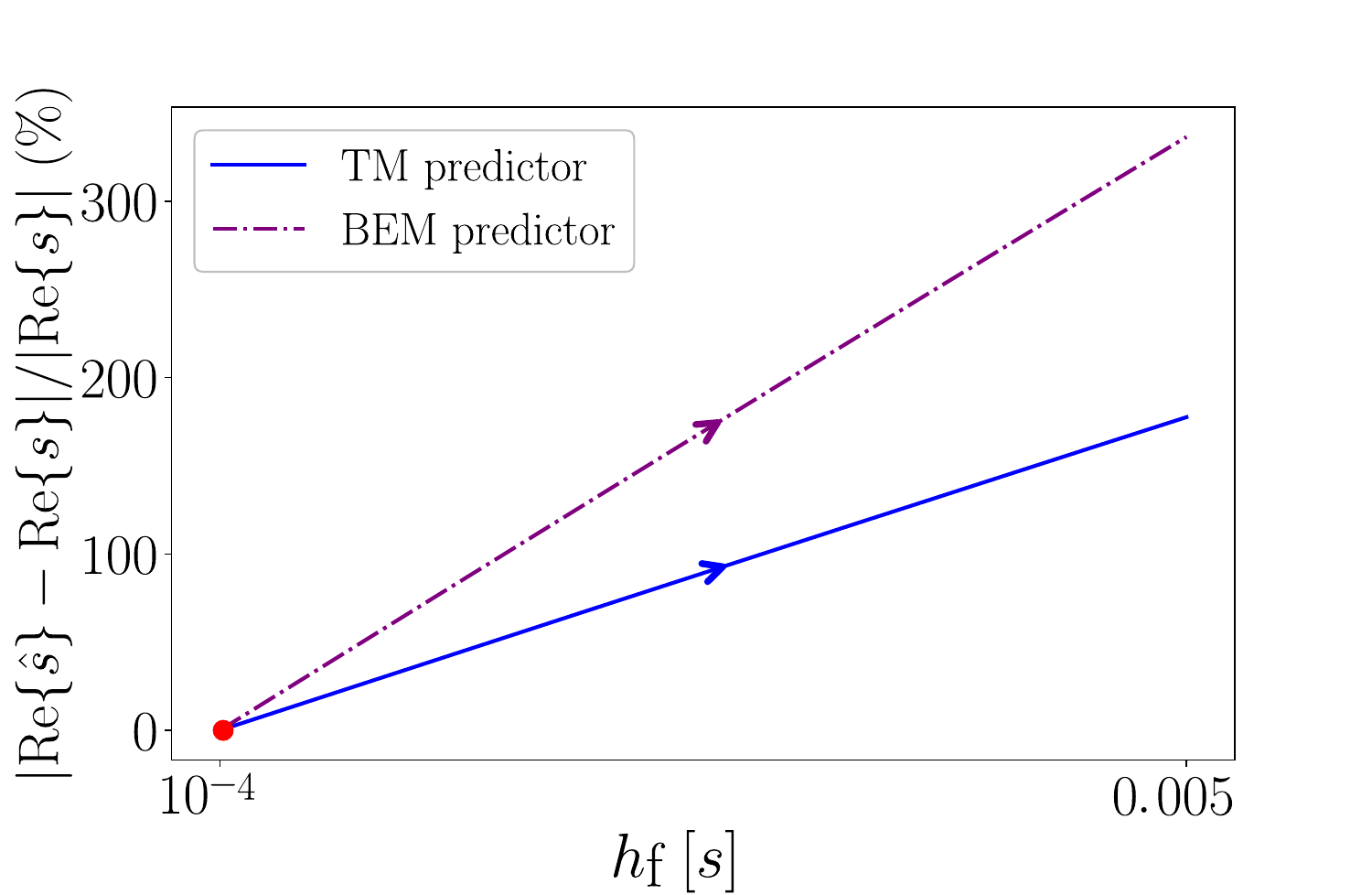}
\caption{Implicit~prediction, BEM solution.}
    \label{fig:c_re}
    \end{subfigure}
    \caption{%\color{blue1}
    {Real part deformation of dominant mode as $h_\ft$ increases from $10^{-4}$ to 0.005~s, $r = 10$.}}
    \label{fig:all_plots_re}
\end{figure*}

\begin{figure*}[t!]
    \centering
    \begin{subfigure}{0.325\textwidth}
        \centering
      \includegraphics[width=\linewidth]{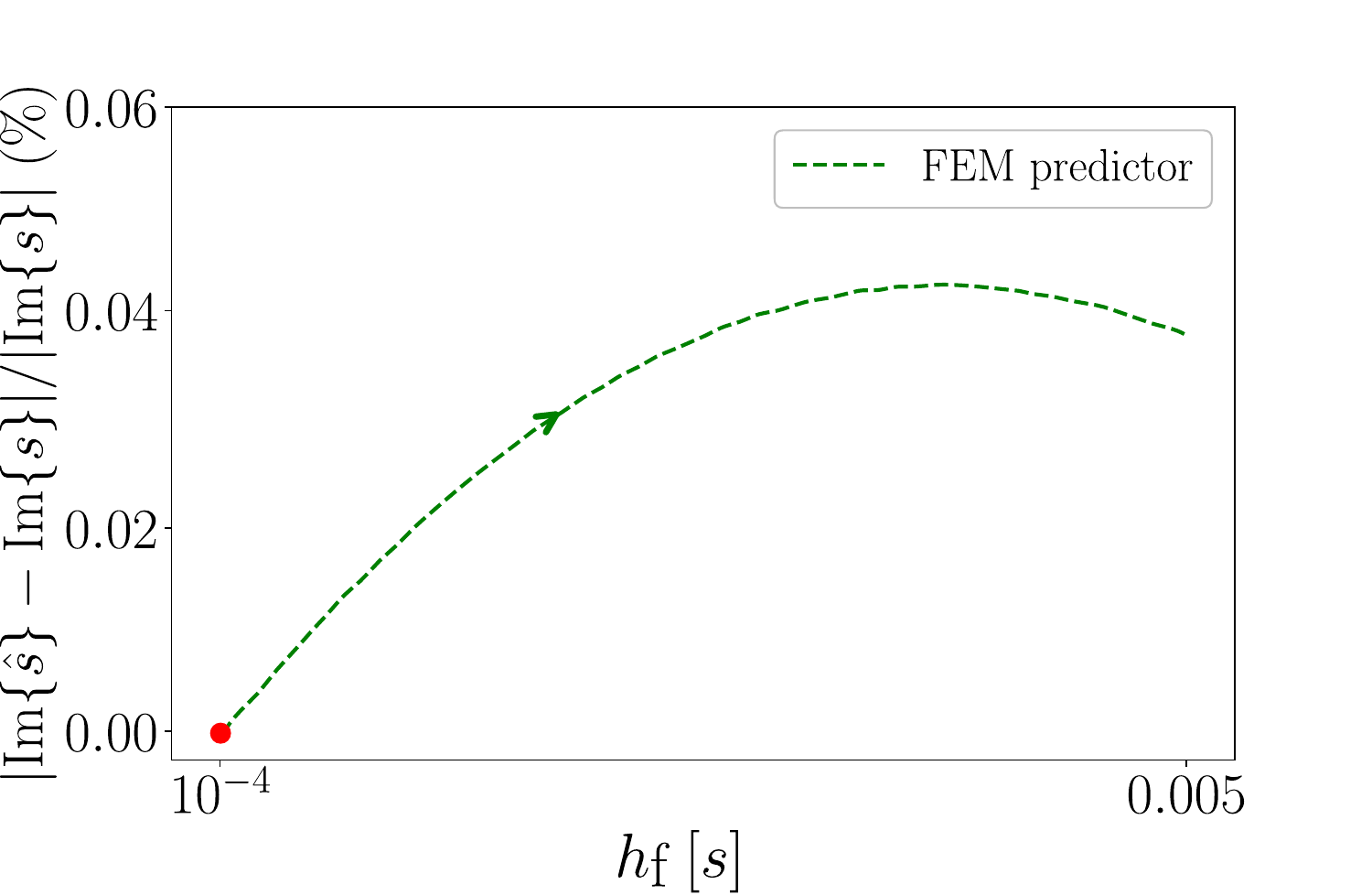}
        \caption{FEM prediction, TM solution.}
        \label{fig:a_im}
    \end{subfigure}
    \hfill
    \begin{subfigure}{0.325\textwidth}
        \centering
    \includegraphics[width=\linewidth]{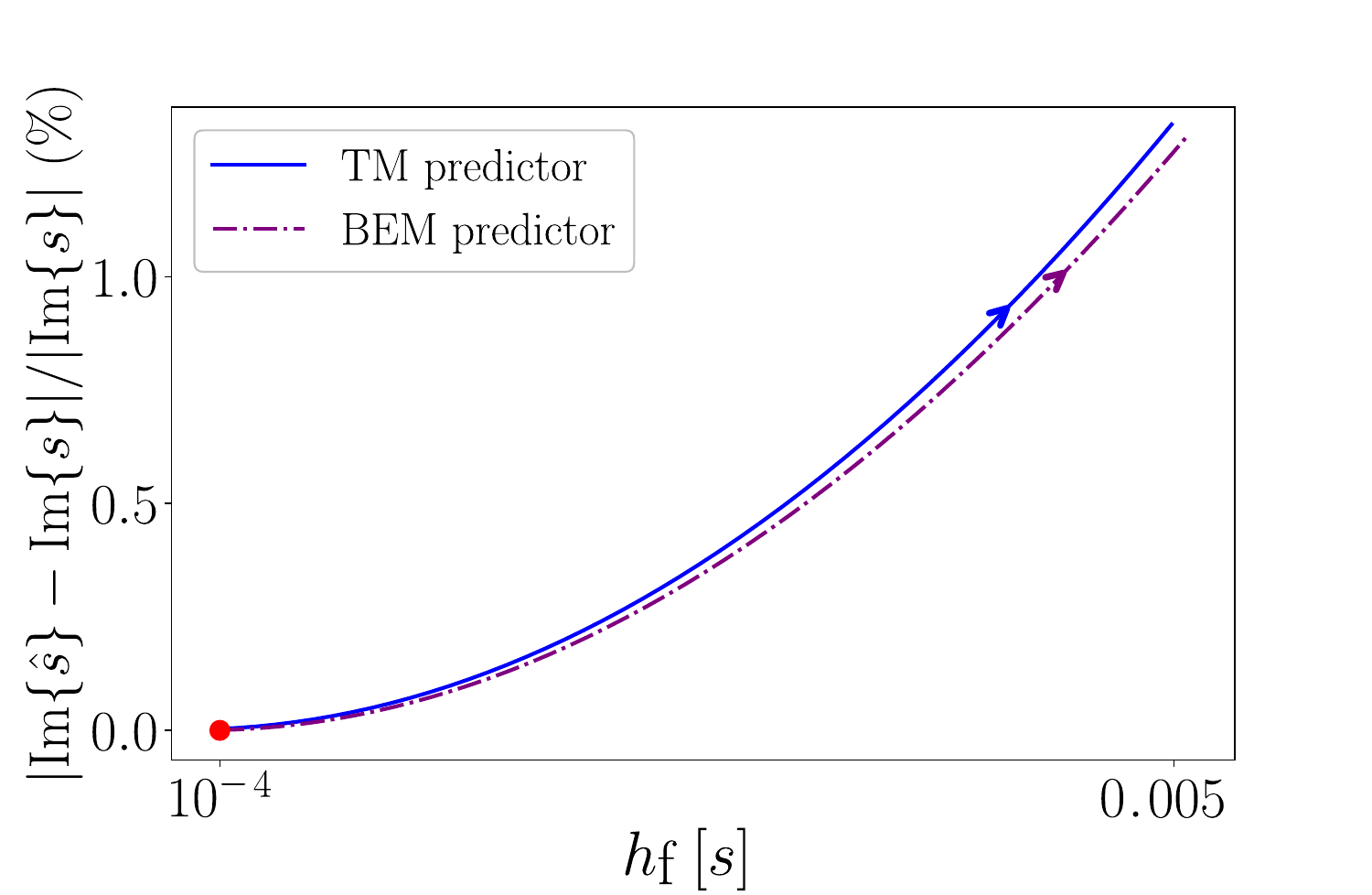}
        \caption{Implicit prediction, TM solution.}
        \label{fig:b_im}
    \end{subfigure}
    \hfill
    \begin{subfigure}{0.325\textwidth}
        \centering
    \includegraphics[width=\linewidth]{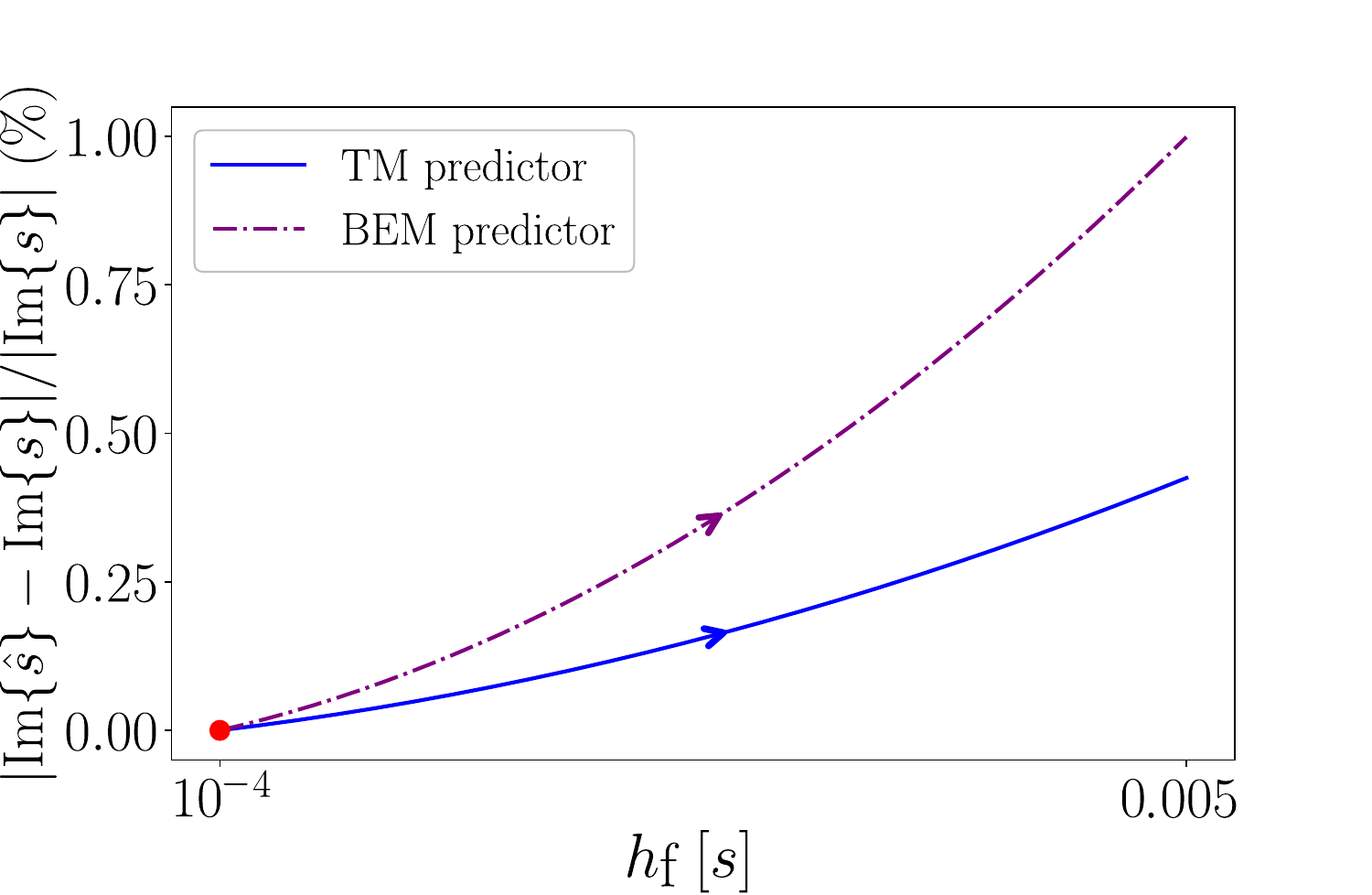}
\caption{Implicit~prediction, BEM solution.}
        \label{fig:c_im}
    \end{subfigure}
    \caption{%\color{blue1}
    {Imaginary part deformation of dominant mode as $h_\ft$ increases from  $10^{-4}$ to 0.005~s, $r = 10$.  }}
    \label{fig:all_plots_im}
\end{figure*}

 We start by partitioning the system variables using the participation matrices $\bfb{P}_x$ and $\bfb{P}_y$ as discussed in Section~\ref{sec:pf} and where we have set $\delta=20$. As a result, the states representing the \ac{avr} voltages, along with the algebraic variables representing the \ac{sg} field voltages, are allocated to the fast subsystem.  For the sake of comparison, we also consider the scenario where state variables are partitioned through standard \ac{pf} analysis, whereas all algebraic variables are classified as fast.  Assuming the multirate scheme described in Subsection~\ref{sec:multirate:illu} -- i.e.,~prediction in step~1) is made with \ac{fem} and solution of fast/slow variables in steps~3) and 4) is achieved with \ac{tm}
-- a first comparison of the two partitioning strategies is provided in Table~\ref{Perror}.  In particular, the table employs the proposed matrix pencil-based approach described in Subsection~\ref{sec:nsssa} to quantify the relative numerical deformation introduced to the system's dominant dynamic mode by the two partitioning strategies.  The dominant mode refers to the local electromechanical oscillation of the \ac{sg} connected to bus~2. In the eigenvalue analysis, this mode is represented by the complex pair $-0.19561 \pm j8.37291$ and has damping ratio  $\zeta = 2.79\%$.  Table~\ref{Perror} suggests that a principled allocation of a subset of the algebraic variables to the slow timescale based on their \acp{pf} achieves accuracy comparable to assuming that all algebraic variables are fast $(\bfg y = \bfg y_{\ft})$, thereby showing potential to improve simulation speed. 
All results in the remainder of this section are produced by considering \ac{pf}-based partitioning of both state and algebraic variables.

%we also establish an empirical partitioning of the system, that is,  %based on our knowledge of its dynamics.  In the empirical partitioning, the dynamics of the stators of the \acp{sg} as well as the states of the \acp{avr}
%are classified as fast, whereas the rest of the system states are classified as slow. Moreover, 
%all algebraic variables are classified as fast. 
%From a practical perspective, the dynamics of the stator and electronic components are relatively fast compared to the slower electrical and mechanical dynamics. Thus, an empirical partitioning can be established by classifying the variables associated with the stator side and the AVR as fast variables, while classifying the remaining variables as slow variables. 
 
We proceed to compare the impact of altering the multirate scheme's predictor method on the numerical deformation introduced to the system dynamic modes.  The results under different time steps for prediction by \ac{fem} and \ac{tm} are shown in Fig.~\ref{fig:predictor}.
As expected, \ac{tm} prediction outperforms \ac{fem} in terms of accuracy.  In fact, as the time steps $h_{\sw}$ and $h_{\ft}$ are increased, the accuracy of \ac{fem} significantly deteriorates and eventually numerical instability is encountered.  On the other hand, with \ac{tm} prediction numerical stability is maintained regardless of the selected time step sizes, although this naturally comes at an additional computational cost due to the need for full Jacobian factorization in step~1).

In Fig.~\ref{fig:all_plots}, we fix the ratio $r=h_{\sw}/h_{\ft}$
to $10$ and study the trajectory of the dominant mode's numerical deformation over the range $h_\ft=[0.0001,0.005]$~s and $h_\sw=[0.001,0.05]$~s. To this end, we consider different implementations of steps~1), 3) and 4) in Section~\ref{sec:multirate:illu}.
In the figure, solving fast/slow variables with \ac{tm} in steps~3) and 4) is referred to as ``\ac{tm} solution".  As shown in Fig.~\ref{fig:a}, the critical value of $h_{\ft}$ beyond which the integration scheme is guaranteed to be destabilized under \ac{fem} prediction is $0.005$~s. 
Moreover, from Fig.~\ref{fig:b} it can be observed that when all steps~1), 3) and 4) are implemented using \ac{tm}, the multirate scheme can exhibit slight underdamping.  This underdamping can be mitigated by using \ac{bem} for prediction.  Yet, as expected, if all steps 1), 3) and 4) are implemented with \ac{bem}, the multirate scheme can lead to significant numerical overdamping.  This is illustrated in Fig.~\ref{fig:c}.
%\color{blue1}  
For the same mode, the numerical deformation of the real and imaginary parts of the corresponding eigenvalue are shown in Figs.~\ref{fig:all_plots_re} and \ref{fig:all_plots_im}, respectively. 
%The results in Fig.~\ref{fig:all_plots_re} confirm the conclusions above regarding damping.  The imaginary part deformation in Fig.~\ref{fig:all_plots_im} highlights the spurious numerical effect on the oscillation frequency of the dominant mode. 
Figure~\ref{fig:all_plots_im} explicitly shows the deformation in the imaginary part, capturing the spurious numerical shift in oscillation frequency.
Under \ac{tm} solution, prediction with \ac{tm} results in slightly greater imaginary part deformation than \ac{bem}. Conversely, under \ac{bem} solution, prediction with \ac{bem} yields smaller deformation than \ac{tm}. 
%It can be concluded that the numerical methods use introduce larger deviations in the real parts compared to the imaginary parts. 
\color{black} 
Overall, the highest level of accuracy in the examined cases is achieved when a \ac{bem} predictor is combined with \ac{tm} solution of fast/slow variables. 

\begin{figure}
\begin{center}
\includegraphics[width=0.55\linewidth]{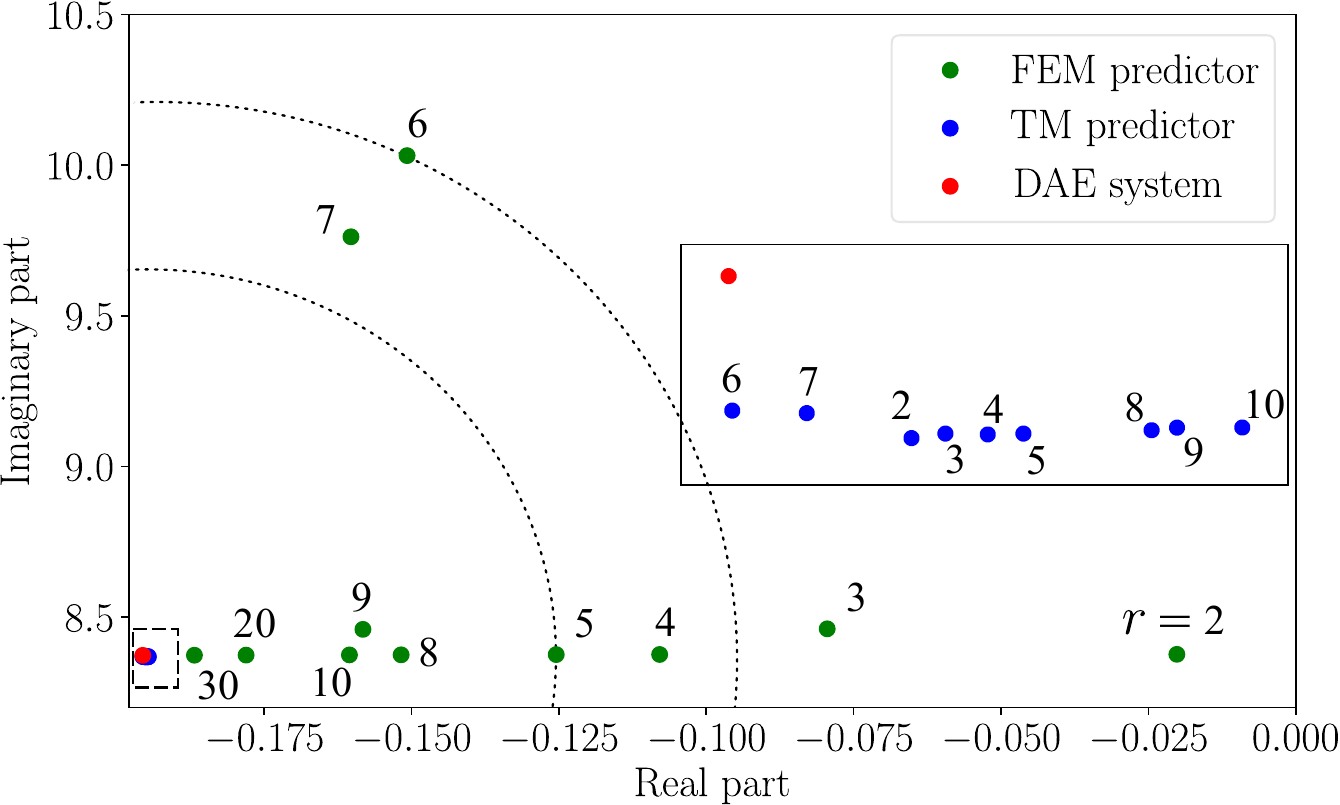} 
\caption{Dominant mode deformation with \ac{fem} and \ac{tm} predictors under different time step ratios ($h_\sw = 0.05$~s). The numbers next to the points represent the ratio values.}
\label{fig:hs}  
\end{center}  
\end{figure}

We further assess the numerical 
deformation of the system's dominant mode by examining the impact of varying the ratio $r=h_\sw/h_\ft$.  To this end, we keep $h_\sw$ constant and vary $r$ by changing $h_\ft$. Fast/slow variables are solved with \ac{tm}.  The results are presented in Fig.~\ref{fig:hs}, and indicate that, interestingly, smaller values of $h_\ft$ do not necessarily imply better accuracy. For example, with \ac{fem} prediction, increasing $r$ from $5$ to $6$ leads to a larger numerical deformation.  Although counterintuitive at a first glance, this irregular behavior occurs as decreasing $h_{\ft}$ increases the number of intermediate steps that rely on linearly interpolated values of the slow variables, which are not updated until $t+h_{\sw}$, thus amplifying interfacing error. 
% \begin{figure}[ht!]
% \begin{center}
% \includegraphics[width=0.55\linewidth]{figs/truncation_error_delta.pdf}    
% \caption{\textcolor{blue1}{Comparison of LTE under three-phase fault with TM-TM under different values of $\delta$.}}
% \label{fig:truncation_error_delta} 
% \end{center}  
% \end{figure}

\begin{figure}
\begin{center}
\includegraphics[width=0.55\linewidth]{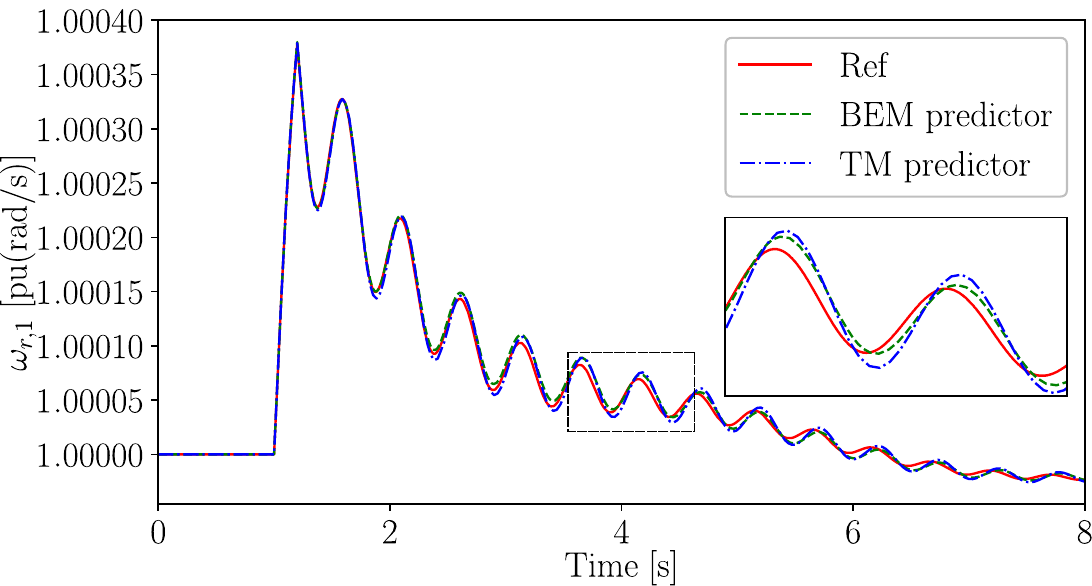}    
\caption{$\omega_{r,1}$ after the load trip at bus~5 with TM solution. $h_\sw = 0.05$~s, $h_\ft = 0.005$~s.}
\label{fig:wscc_load_tm}  
\end{center}  
\end{figure}

\begin{figure}
\begin{center}
\includegraphics[width=0.55\linewidth]{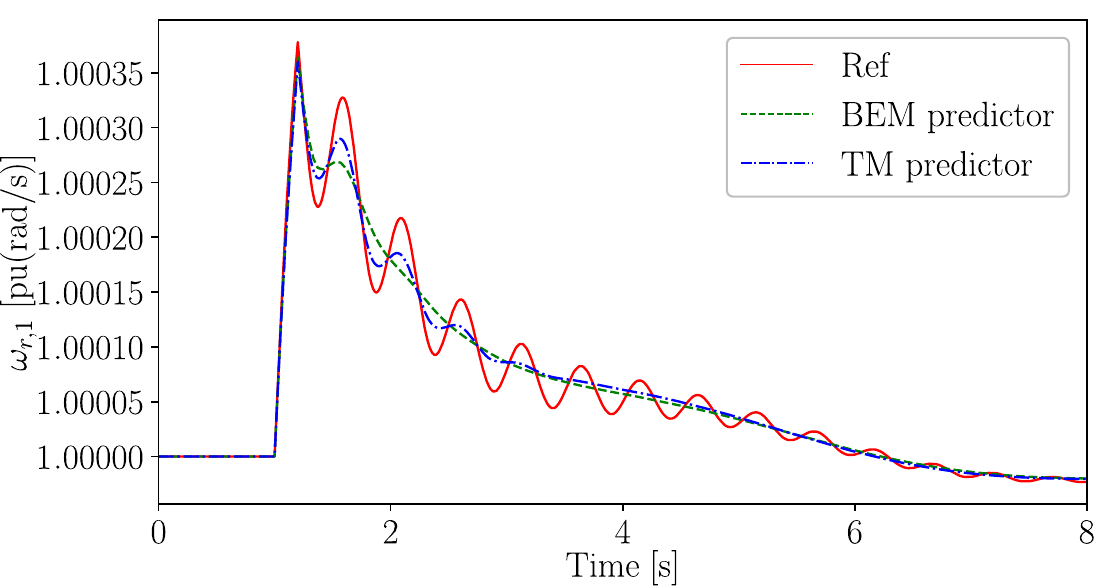}    
\caption{$\omega_{r,1}$ after the load trip at bus 5 with BEM solution. $h_\sw = 0.05$~s, $h_\ft = 0.005$~s.}
\label{fig:wscc_load_bem}  
\end{center}  
\end{figure}

\begin{figure}[ht!]
\begin{center}
\includegraphics[width=0.55\linewidth]{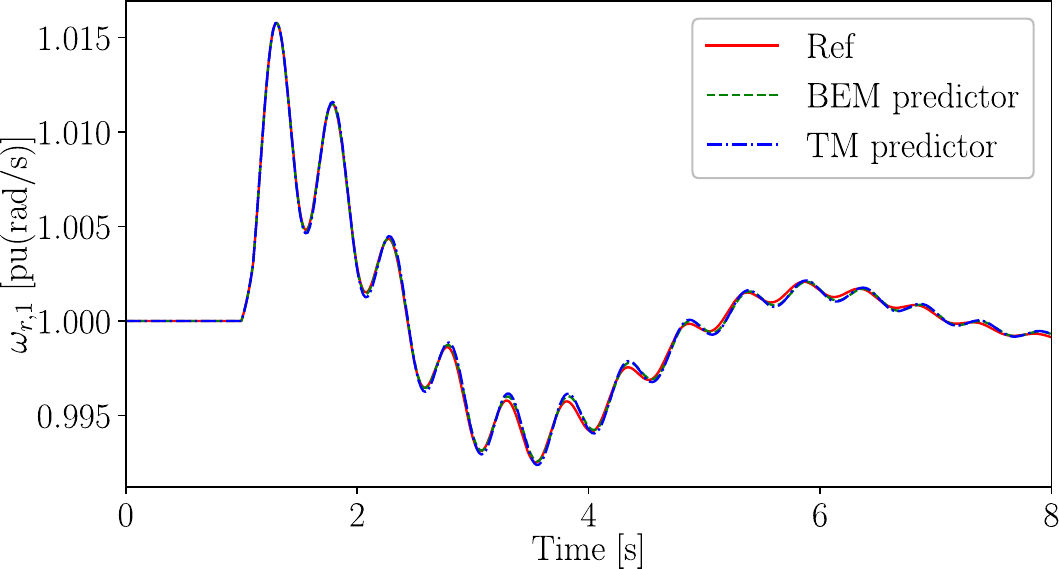}    
\caption{$\omega_{r,1}$ after the fault at bus~5 with TM solution. $h_\sw = 0.05$~s, $h_\ft = 0.005$~s.}
\label{fig:wscc_short_tm} 
\end{center}  
\end{figure}

\begin{figure}[ht!]
\begin{center}
\includegraphics[width=0.55\linewidth]{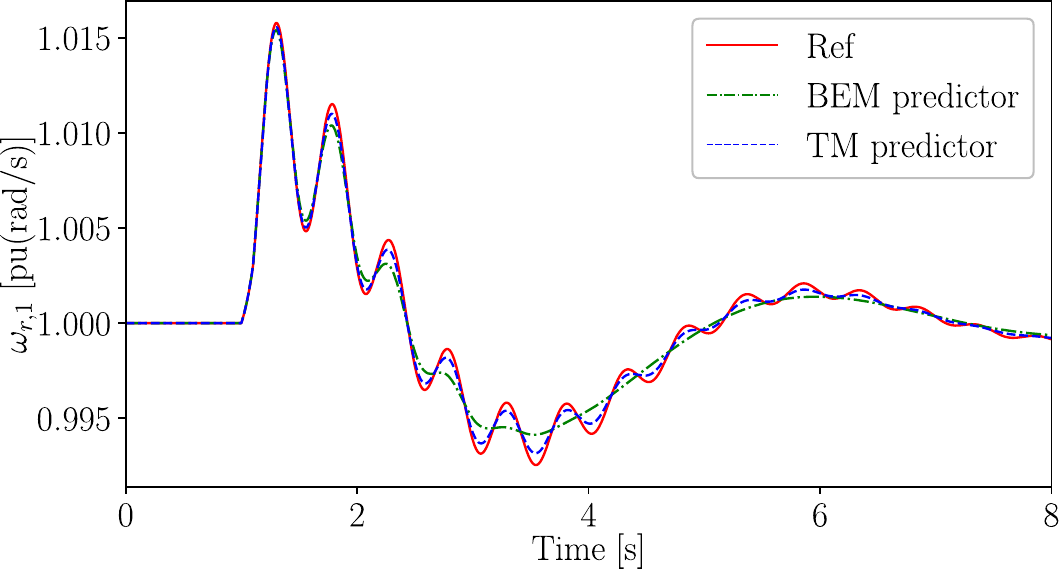}    
\caption{$\omega_{r,1}$ after the fault at bus~5 with BEM solution. $h_\sw = 0.05$~s, $h_\ft = 0.005$~s.}
\label{fig:wscc_short_bem}  
\end{center}  
\end{figure}

% \begin{figure}[ht!]
% \begin{center}
% \includegraphics[width=0.55\linewidth]{figs/truncation_error_single_multi.pdf}    
% \caption{\textcolor{blue1}{Comparison of LTE under three-phase fault using different approaches: single-rate and multirate.}}
% \label{fig:truncation_error_single_multi}  
% \end{center}  
% \end{figure}

In the following, we perform multirate non-linear time-domain simulations of the system's non-linear \ac{dae} model.  %\color{blue1} 
%to further check the results obtained through \ac{sssa}. 
%\color{black}  
In particular, we consider the response of the system after (i) a sudden trip of 50\% of the load connected to bus~5 occurring at $ t = 1$~s and followed by reconnection at $ t = 1.2$~s; and (ii) a three-phase fault occurring at bus~5 at $t=1$~s and cleared after $100$~ms by tripping the line that connects buses 5 and 7.  The simulation results in Fig.~\ref{fig:wscc_load_tm} to Fig.~\ref{fig:wscc_short_bem} show the response of $\omega_{{\rm r},1}$ following each disturbance.  Four different multirate setups are considered for prediction of slow variables and solution of fast/slow variables, namely, \ac{bem} prediction with \ac{tm} solution; \ac{tm} prediction with \ac{tm} solution;
\ac{bem} prediction with \ac{bem} solution; and \ac{tm} prediction with \ac{bem} solution.  
%\color{blue1}  
Reference trajectories here and in the remainder of the paper are obtained using a two-stage diagonally implicit Runge-Kutta method with time step $h=0.001$~s.
%\color{black}  
From Fig.~\ref{fig:wscc_load_tm} and Fig.~\ref{fig:wscc_short_tm}, it can be seen that \ac{tm} solution can lead to slight underdamping, which is mitigated when coupled with \ac{bem} prediction.  Moreover, from Fig.~\ref{fig:wscc_load_bem} and Fig.~\ref{fig:wscc_short_bem}, it can be seen that \ac{bem} solution 
%\color{blue1}
can result in 
%\color{black} 
significant numerical overdamping, which is mitigated when combined with \ac{tm} predictor. 
%\color{blue1}
The results also indicate that the multirate scheme introduces small oscillation frequency shifts.  
These results are consistent with the conclusions obtained from the matrix pencil-based \ac{sssa} in 
%\color{blue1}
Figs.~\ref{fig:all_plots}-\ref{fig:all_plots_im}. 
%a slight shift in oscillation frequency due to the imaginary part deformation.  However, since the deformation in the imaginary part remains small, less than $1.05~\%$ for the current time step used, as seen in Figs.~\ref{fig:b_im} and~\ref{fig:c_im}), the resulting phase deviation is less than $5^\circ$.  

\begin{figure}[ht!]
    \centering
    \begin{subfigure}[b]{0.48\linewidth}
        \includegraphics[width=\textwidth]{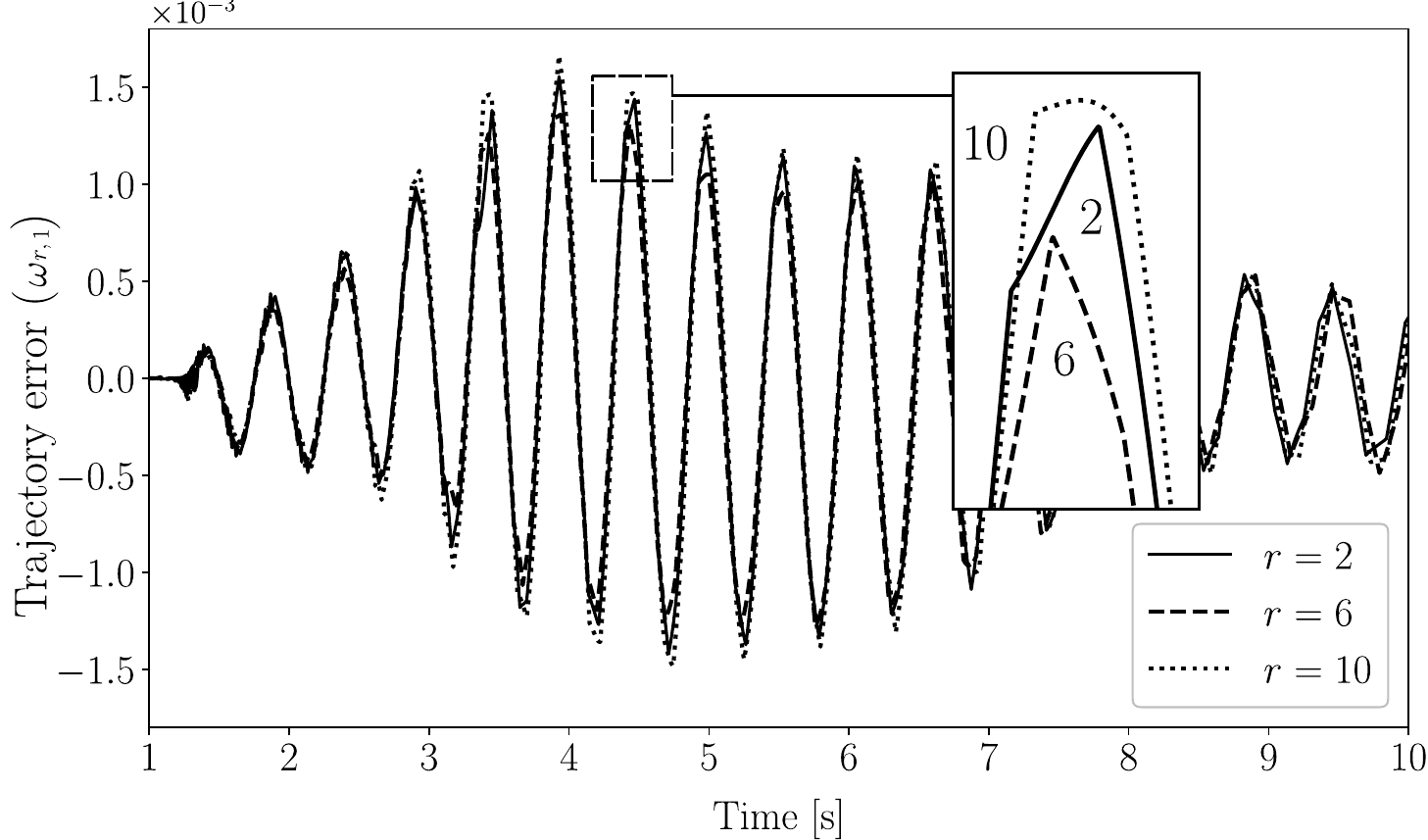}
        \caption{Linear interpolation.}
        \label{fig:LTE_r_linear}
    \end{subfigure}
    \hfill  
    \begin{subfigure}[b]{0.48\linewidth}
    \includegraphics[width=\textwidth]{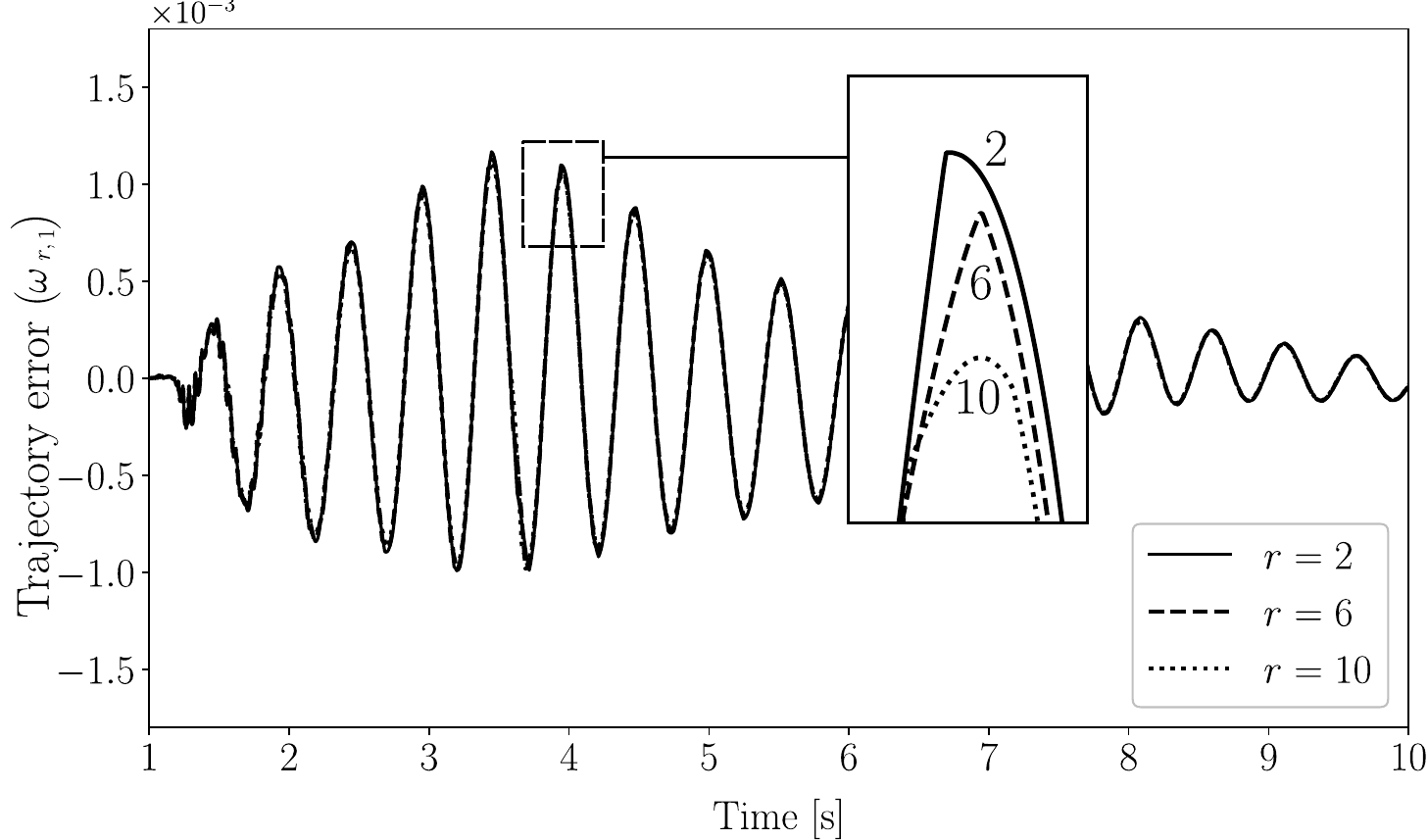}
    \caption{Cubic spline interpolation.}
    \label{fig:LTE_r_spline}
    \end{subfigure}    
    \caption{
    %\color{blue1}
    {Trajectory error of $\omega_{{\rm r},1}$ under fault at bus~5: linear vs. cubic spline interpolation.}}
    \label{fig:LTE_r}
\end{figure}

We further examine the effect of interpolation of the slow variables.
Figure~\ref{fig:LTE_r} shows the numerical error of the trajectory of machine~1 ($\omega_{{\rm r},1}$) 
following the fault at bus~5, comparing linear and cubic spline interpolation for different time step ratios $r$, using \ac{tm} prediction and \ac{tm} solution.  Figure~\ref{fig:LTE_r_linear} confirms the occurrence of irregular numerical deformation
when linear interpolation is used (see also Fig.~\ref{fig:hs}).  In contrast, these irregularities are eliminated when cubic spline interpolation is applied, as shown in Fig.~\ref{fig:LTE_r_spline}. In all cases examined, spline interpolation yields smaller errors than linear interpolation.
%, while Fig.~\ref{fig:LTE_r_spline} does not exhibit the same pattern, instead, the error is decreasing with ratio increases. Overall, the spline interpolation yields smaller error compared to the linear method.
\color{black}

 \begin{figure*}[ht!]
    \centering
    \begin{subfigure}{0.325\textwidth}
        \centering
    \includegraphics[width=\linewidth]{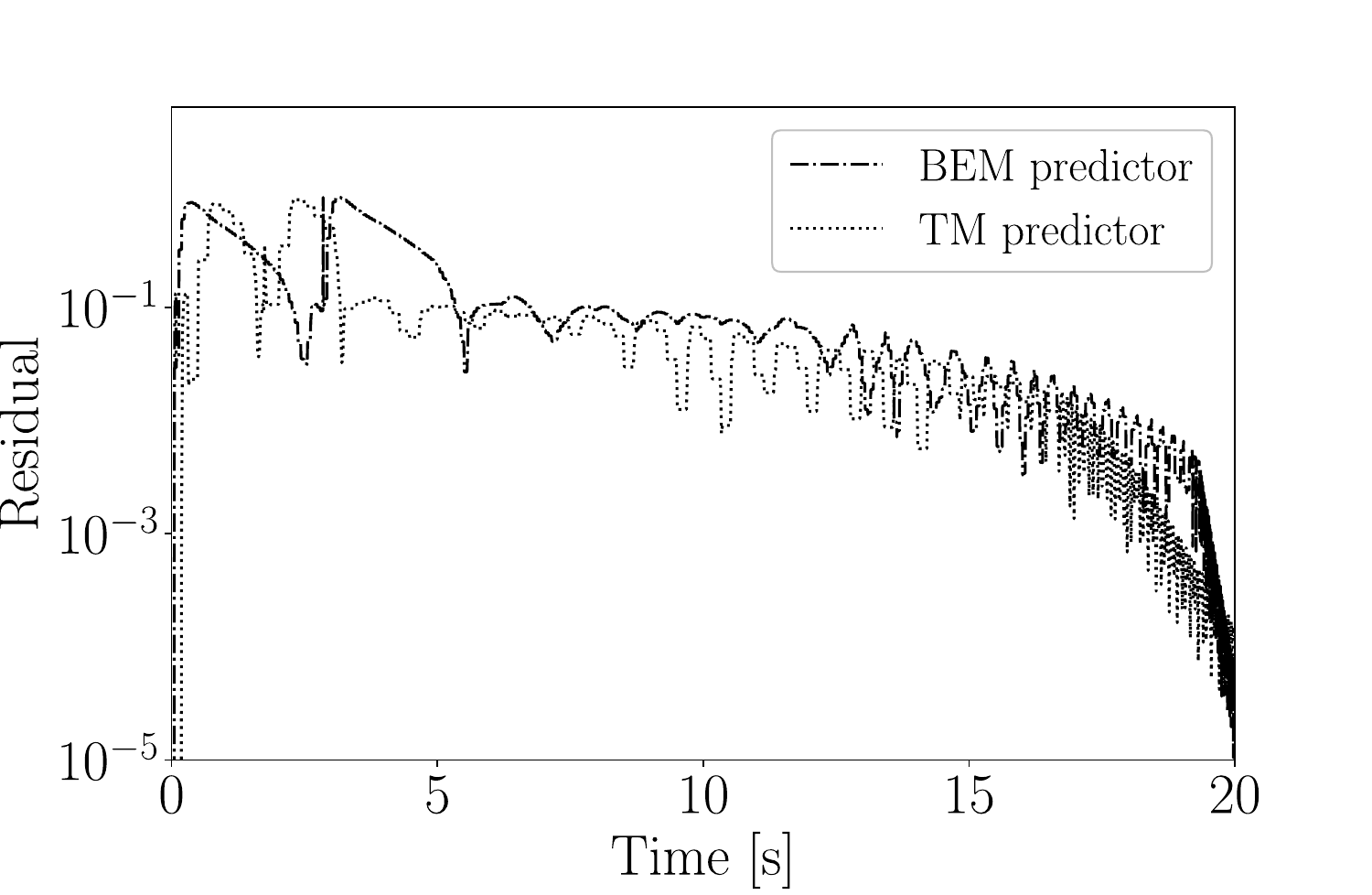}
        \caption{Predictor: BEM vs. TM.}
        \label{fig:residul_pre}
    \end{subfigure}
    \hfill
    \begin{subfigure}{0.325\textwidth}
        \centering
\includegraphics[width=\linewidth]{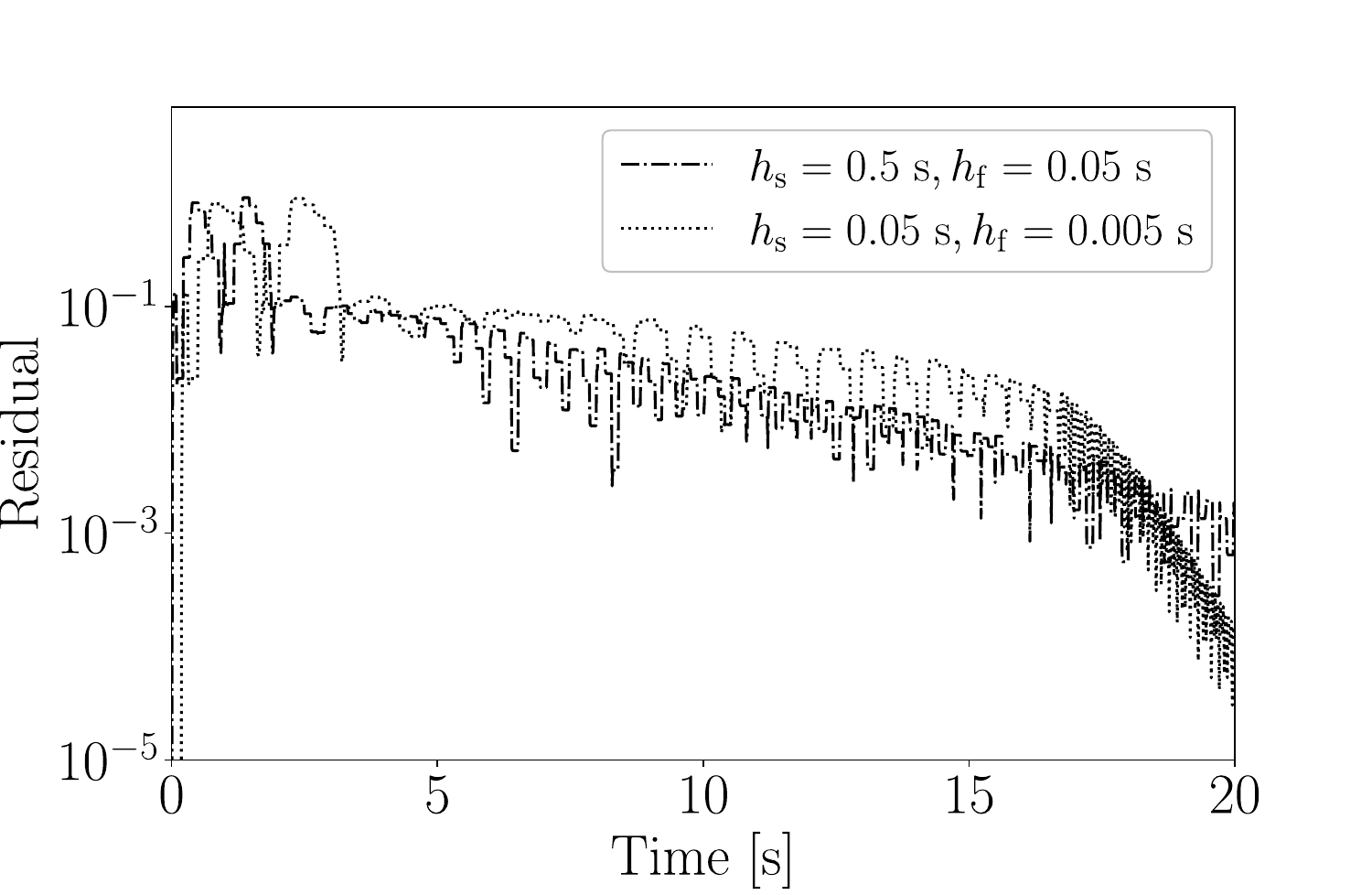}
        \caption{Different time steps.}
        \label{fig:res_time}
    \end{subfigure}
    \hfill
    \begin{subfigure}{0.325\textwidth}
        \centering
    \includegraphics[width=\linewidth]{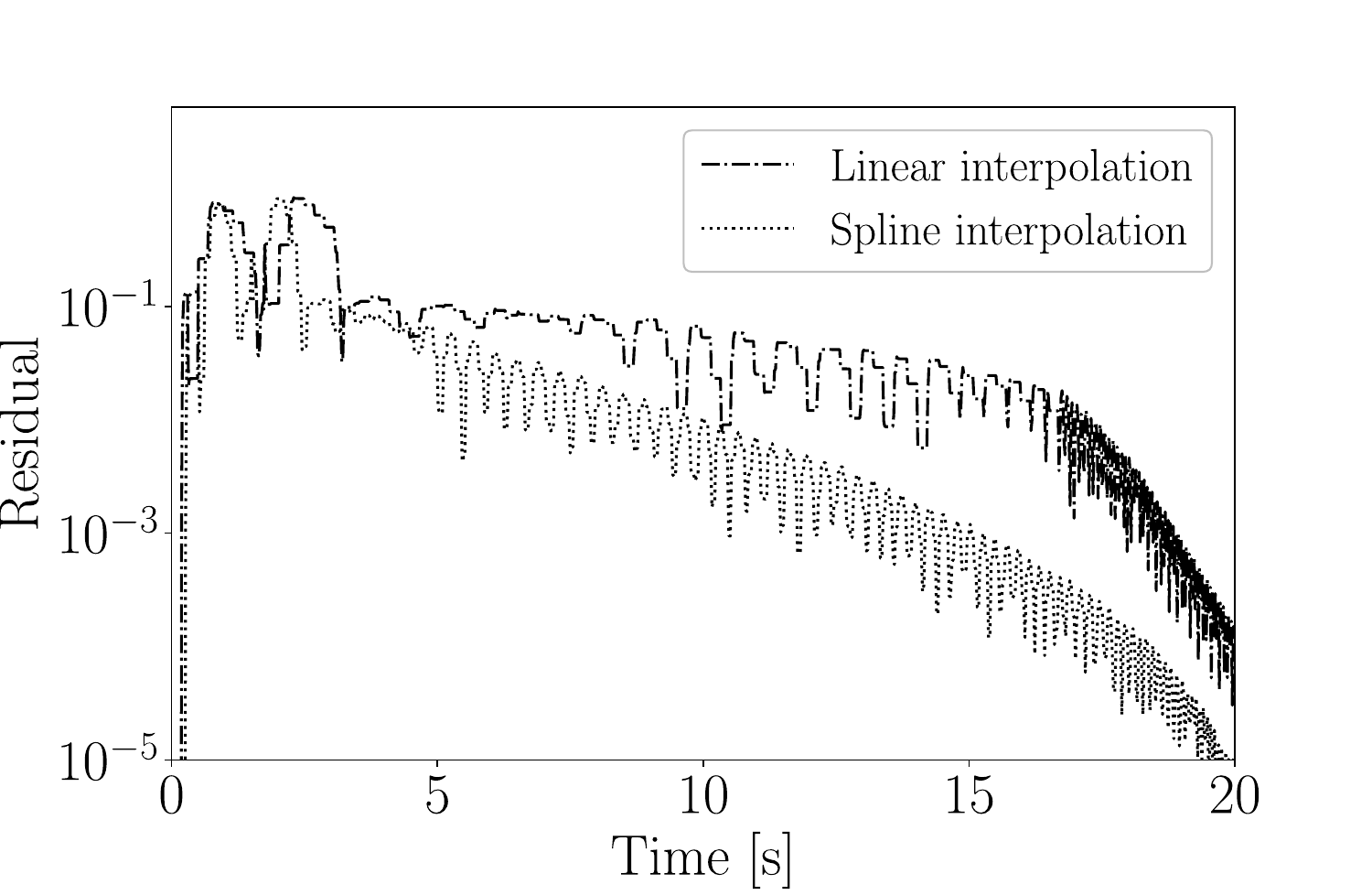}
        \caption{Interpolation: linear vs. cubic spline.}
        \label{fig:res_inter}
    \end{subfigure}
    \caption{
    %\color{blue1}
    {Fault at bus~5, TM solution ($h_s = 0.005$~s, $h_f = 0.001$~s): algebraic residual of slow subsystem.}}
    \label{fig:residual}
\end{figure*}

%\color{blue1} 
Another relevant issue associated with interpolation of the slow variables is the appearance of algebraic constraint residuals.  As described in Section~\ref{sec:multirate:form}, fast variables are solved at every step, while the slow variables are updated at the slow time scale. During intermediate steps, the slow variables are not solved but interpolated, so the algebraic equations involving $\bfg y_{\rm s}$ may indeed exhibit residuals.
Figure~\ref{fig:residual} shows the evolution of the Euclidean norm of the algebraic constraint residuals of the slow subsystem.  The residuals increase following the disturbance and gradually decrease as the system approaches steady-state conditions.  Yet, it is worth noting that even if residuals are very small, this does not imply that the overall numerical solution is accurate.

%It should be noted that residual growth during transients does not imply any numerical instability. Our results confirm that the simulation preserves numerical stability even when residuals temporarily increase.
\color{black}

%During the simulation, fast variables are solved at every step while the slow variables are updated at slow time scale as described in Section~\ref{sec:multirate:form}. During intermediate steps, the slow variables are not solved but interpolated, so the algebraic equations involving $\bfg y_{\rm s}$ may indeed exhibit residuals. 

%\color{blue1}  
%Next, we will show the impact of different thresholds and ratios on simulation errors.  The error showed below denotes the trajectory error of the response of the rotor speed of machine 1 ($\omega_{{\rm r},1}$) during the simulation compared with the reference results.  
%\color{black}

\begin{figure*}[ht!]
    \centering
    \begin{subfigure}{0.325\textwidth}
        \centering 
 \includegraphics[width=\linewidth]{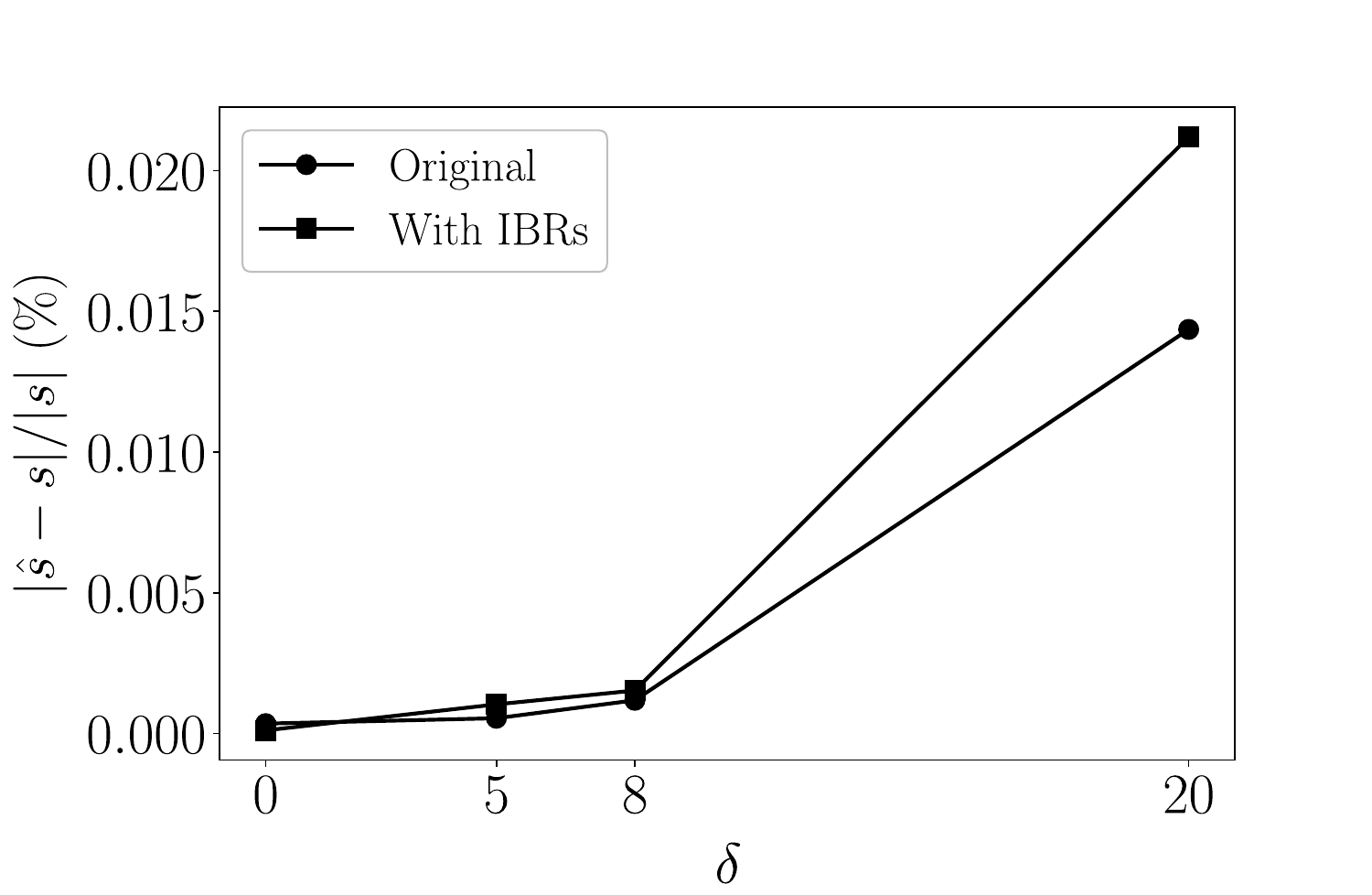}
        \caption{Dominant mode numerical deformation.}
        %TM prediction, TM solution.}
        \label{fig:tmtm}
    \end{subfigure}
    \hfill
    \begin{subfigure}{0.325\textwidth}
    \centering
    \includegraphics[width=\linewidth]{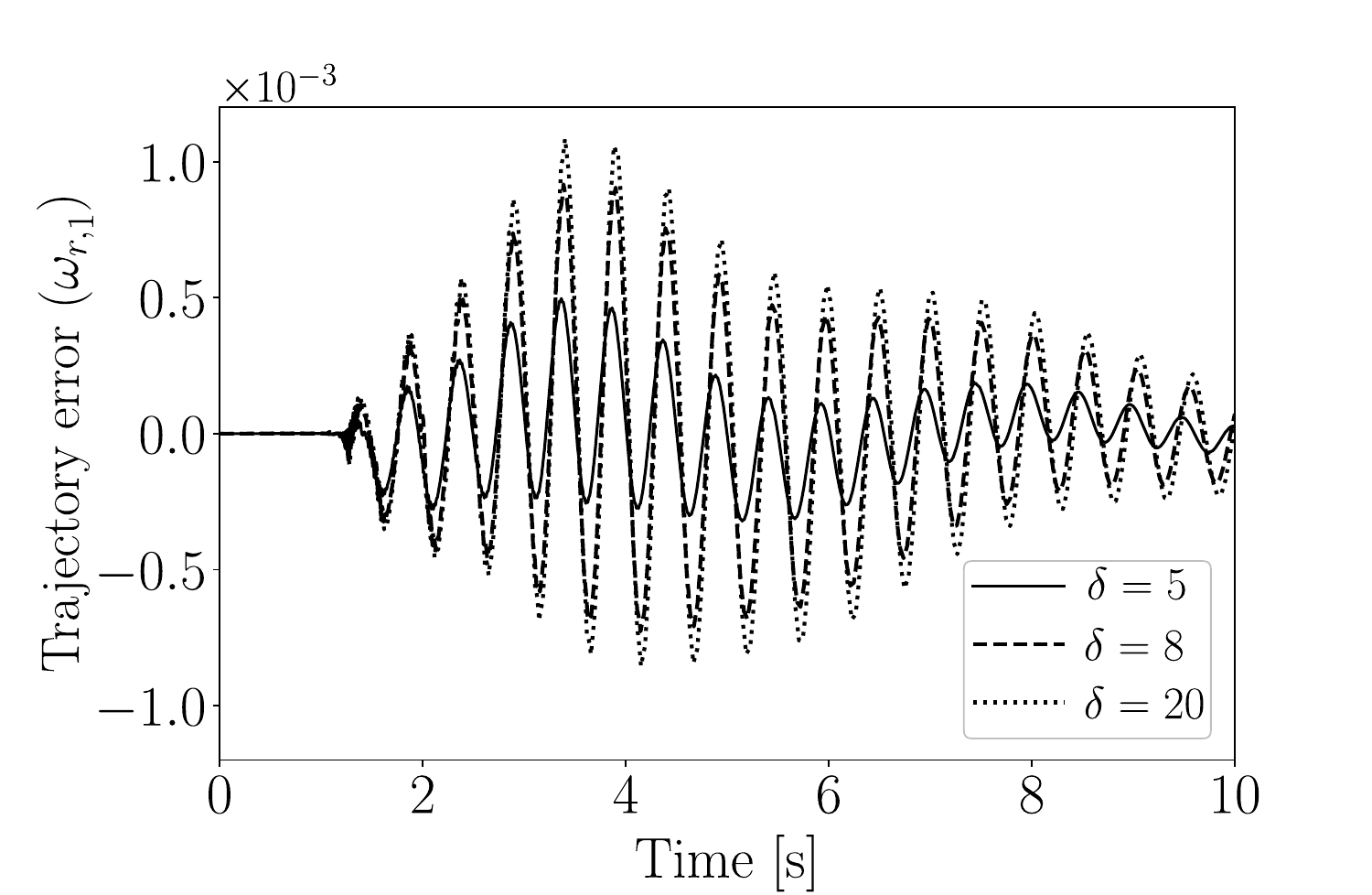}
    \caption{Original system.}
    \label{fig:original}
    \end{subfigure}
    \hfill
    \begin{subfigure}{0.325\textwidth}
        \centering
    \includegraphics[width=\linewidth]{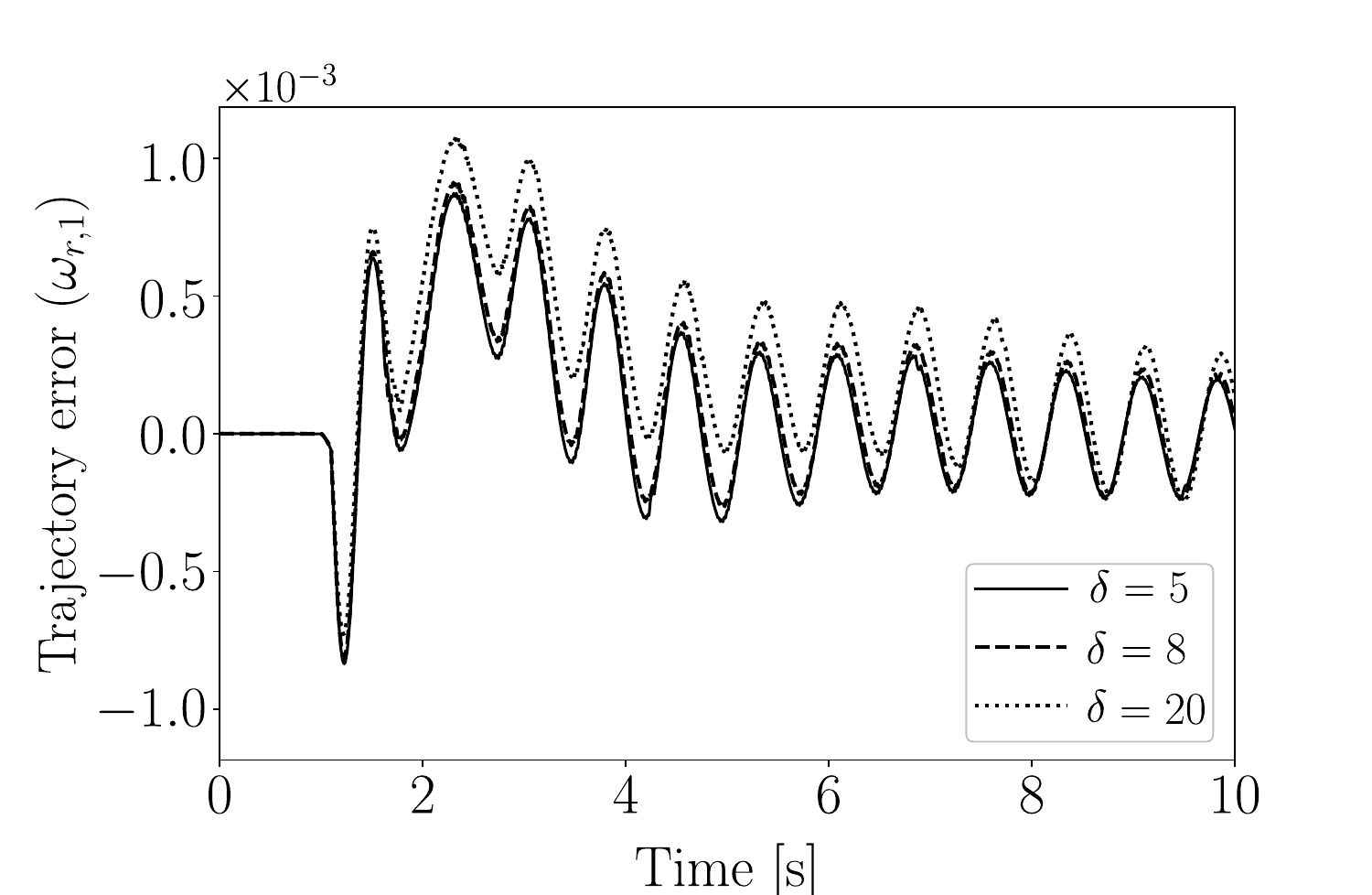}
        \caption{Modified system with IBRs.}
        \label{fig:der}
    \end{subfigure}
    \caption{%\color{blue1}
    {Impact of changing the partitioning threshold $\delta$ ($h_s = 0.005$~s, $h_f = 0.001$~s).}}
    \label{fig:eig_error}
\end{figure*}

%\color{blue1}
%For our partitioning strategy, the threshold $\delta$ for distinguishing fast and slow variables should  be firstly determined based on system modes (i.e., the magnitude of eigenvalues). 
We finally discuss the impact of varying the partitioning threshold $\delta$ on mode deformation for both the original WSCC system and a modified version with increased stiffness, where part of the synchronous generation is replaced by \acp{ibr}.  As seen in Fig.~\ref{fig:tmtm}, for a given fast time step size, $\delta=0$ leads to a single-rate simulation where all variables are integrated with the fast step, resulting in minimal mode deformation for this setup.  As $\delta$ increases, more variables are assigned to the slow subsystem, and for a fixed $h_{\sw}$, the deformation tends to grow.
We note that, in practice, varying 
$\delta$ would generally affect the selection of $h_{\sw}$ to provide a good trade-off between accuracy and computational cost.
%to capture the dynamics of the variables included in the slow subsystem.
Figures~\ref{fig:original} and \ref{fig:der} compare the error during the three-phase fault at bus~5 using TM prediction with TM solution for different values of $\delta$; the results are consistent with those of small-signal analysis.

%\color{blue1}
%Figure~\ref{fig:original} compares the error during three-phase faults using the TM prediction with TM solution for different $\delta$, the results are consistent with the \ac{sssa} result presented in Fig.~\ref{fig:tmtm}. We also apply the same three-phase fault to the modified WSCC system and show the error with different $\delta$ in Fig.~\ref{fig:der}. %The results further verify that the error is increasing when $ \delta$ increases. 
\color{black}

%While $\delta=20$ exhibits the largest deformation among the tested $\delta$, the resulting error remains within the acceptable range (less than 0.02\%).
\color{black}

\section{Conclusion}
\label{sec:conclusion}

The paper presents a matrix pencil-based approach for analyzing the numerical stability and accuracy of multirate time-domain simulation schemes applied to power system \acp{dae}.  By expressing the discretized small-signal multirate model as a system of linear difference equations, our approach enables the systematic assessment of spurious numerical mode deformation.
In addition, a principled strategy for partitioning both state and algebraic variables based on modal participation factors is discussed.
The proposed approach is used to evaluate the impact of different predictor types, integration methods, and time step sizes, revealing nontrivial accuracy trends.  Our findings support the use of small-signal tools to guide the design of multirate schemes in practical applications.

Future work will explore the application of multirate methods in co-simulation frameworks for large-scale power systems involving electromagnetic transients, as well as further investigate the capabilities of PF-based partitioning strategies.

\section{Funding}

This work is supported by the China Scholarship Council, by funding 
L.~Huang; and by Science Foundation Ireland, by funding G.~Tzounas under project NexSys (grant. no.~21/SPP/3756).

\appendix
%\color{blue1}
 \section[Participation matrix of algebraic variables]{Derivation of \eqref{eq:P_y}}
\label{sec:appA}

By treating algebraic variables as outputs of the system's state-space with output matrix $\bfg C \in \mathbb{R}^{m \times n}$, i.e.,~$\wdt{ \bfg y} = \bfg C  \wdt {\bfg x}$,
from the second equation in \eqref{eq:linear} we get that $\bfg C = -\bfg g^{-1}_{y}\bfg g_{x} $.
Denoting the $\mu$-th row as $\bfg C_{\mu} = [c_{\mu1}, \dots, c_{\mu n}]$, $\wdt{y}_{\mu}$ is given by:
%with output matrix $\bfg C \in \mathbb{R}^{m \times n}$, where its $\mu$-th row vector $\bfg C_{\mu} = [c_{\mu1}, \dots, c_{\mu n}]$ characterizes the $\mu$-th output dynamics. Accordingly, the expression for 
%
\begin{equation}
    \wdt{ y}_{\mu} = \bfg {C}_{\mu} \wdt {\bfg x}= c_{\mu 1} \wdt {x}_1 + c_{\mu 2} \wdt {x}_2 + \dots + c_{\mu n} \wdt {x}_n
    \label{eq:ymu}
\end{equation}
And, by using \eqref{eq:xsolution}:
\begin{align}
     \wdt{ y}_{\mu}
   %& \nonumber = c_{\mu 1} \sum_{i=1}^{n} p_{1i} ^{[x]}e^{s_i t} + \cdots + c_{\mu n}  \sum_{i=1}^{n} p_{ni} ^{[x]}e^{s_i t}   \\
& \nonumber =\sum_{i=1}^{n}  (c_{\mu 1} p_{1i} ^{[x]}  + \cdots + c_{\mu n}   p_{ni} ^{[x]} ) e^{s_i t} 
%\\
%   & 
   = \sum_{i=1}^{n} p_{\mu i} ^{[y]} e^{s_i t}
\end{align}
where $ p_{\mu i} ^{[y]}  = c_{\mu 1} p_{1i} ^{[x]} + \cdots + c_{\mu n} p_{ni} ^{[x]}$ is the $(\mu,i)$-th element of the participation matrix of algebraic variables $\mathbf{P}_{y}$ as defined in \eqref{eq:P_y}.

 \color{black}
\section{Proof of Proposition 1}
\label{sec:appB}
 
We start by introducing parameters $a, a^*$ for the predictor used in step 1) and $b, b^*, c, c^*$ for the solution of fast and slow variables in steps 3) and 4), respectively.  These parameters allow the proof to accommodate different combinations of predictor and corrector methods in a unified manner.

For \ac{fem} prediction, $a = h_{\sw}$ and $a^* = 0$; 
for \ac{tm} prediction, $a = a^* = h_{\sw}/2$; 
and for \ac{bem} prediction, $a = 0$ and $a^* = h_{\sw}$. 
For the solution of fast and slow variables with \ac{tm}, $b = b^* = h_{\ft}/2$ and $c = c^* = h_{\sw}/2$; 
with \ac{bem}, $b = c = 0$, $b^* = h_{\ft}$, and $c^* = h_{\sw}$.

%We proceed with the proof based on the illustrative multirate scheme in Section~\ref{sec:multirate:illu}.
%
Incorporating the parameters $a, a^*,b,b^*,c,c^*$,
\eqref{eq:li}, \eqref{li_3} and \eqref{eq:li_4} are rewritten as follows:
%\color{blue1}
%
\begin{equation}
\begin{aligned}
\wdt{\bfg{x}}^P_{t+h_\sw} &= \wdt{\bfg{x}}_{t} + a\left( \bfg{f}_x {\wdt{\bfg{x}}_t} + \bfg{f}_y
\wdt{\bfg{y}}_t \right) +a^* \left( \bfg{f}_x {\wdt{\bfg{x}}_{t+h_\sw}} + \bfg{f}_y
\wdt{\bfg{y}}_{t+h_\sw} \right) \\
\bfg 0_{m,1} &= \bfg{g}_x  \wdt{\bfg{x}}^P_{t+h_\sw} +  \bfg{g}_y \wdt{\bfg{y}}^P_{t+h_\sw}
\end{aligned}
\label{eq:li_a}
\end{equation}
\begin{equation}
\begin{aligned}
%\nonumber
\wdt{\bfg x}_{\ft, t + i h_\ft} =& 
\wdt{\bfg x}_{\ft,{t} + (i - 1)h_\ft} 
  + b^*
  ( \bfg f_{\ft,x}
  \wdt{\bfg x}_{t + i h_\ft} + {\bfg f_{\ft,y}} \wdt{\bfg y}_{{t} + ih_\ft} )
  \\
  \label{li_3_b}
  & 
 \quad  \!\!\!\!\!
  + b
  ( {\bfg f_{\ft,x} \wdt{\bfg x}_{t + (i - 1)h_\ft} \!\! + \! \bfg f_{\ft,y} \wdt{\bfg y}_{t + (i - 1)h_\ft}} ) 
  \\
  \bfg 0_{m_\ft,1} & 
=b^*
\left( \bfg g_{\ft,x}
\wdt{\bfg x}_{t + i h_\ft} + \bfg g_{\ft,y} \wdt{\bfg y}_{t + i h_\ft} 
\right)
%\nonumber
\end{aligned}
\end{equation}
\begin{equation}
\begin{aligned}
\wdt{\bfg x}_{\sw, t + h_\sw} &=
\wdt{\bfg x}_{\sw,t} + c^*
( \bfg f_{\sw,x} \wdt{\bfg x}_{t + h_\sw} + \bfg f_{\sw,y} \wdt{\bfg y}_{t + h_\sw} )
+c
  ( \bfg f_{\sw,x} \wdt{\bfg x}_{t}  +  \bfg f_{\sw,y} \wdt{\bfg y}_{t} ) 
  \\
\bfg 0_{m_\sw,1} & 
= c^*
\left( \bfg g_{\sw,x}
\wdt{\bfg x}_{t + h_\sw} + \bfg g_{\sw,y} \wdt{\bfg y}_{t + h_\sw} 
\right)
\end{aligned}
 \label{eq:li_4_c}
 \end{equation}
 We define: 
\begin{equation}
\bfg f_x =
\begin{bmatrix}
\bfg{f}_{\ft\ft,x} & \bfg{f}_{\ft\sw,x} \\
\bfg{f}_{\sw\ft,x} & \bfg{f}_{\sw\sw,x} 
\end{bmatrix}
\end{equation}
where 
${\bfg{f}_{\ft\ft,x}} \in \mathbb{R}^{{{n_\ft} \times {n_\ft}}}$, ${\bfg{f}_{\ft\sw,x}} \in \mathbb{R}^{{{n_\ft} \times {n_\sw}}}$, ${\bfg{f}_{\sw\ft,x}} \in \mathbb{R}^{{{n_\sw} \times {n_\ft}}}$ and ${\bfg{f}_{\sw\sw,x}} \in \mathbb{R}^{{{n_\sw} \times {n_\sw}}}$. Similarly for $\bfg f_y$, $\bfg g _x$ and $\bfg g_y$. 
Dividing 
\eqref{eq:li_a} into fast and slow timescales, we have: 
\begin{equation}
\begin{aligned}
   \wdt{\bfg x}^P_{\ft,{t} + h_\sw}  &= \wdt{\bfg x}_{\ft,{t}} + a{\bfg f_{\ft\ft,x}} \wdt{\bfg x}_{\ft,{t}} + a{\bfg f_{\ft\sw,x}} \wdt{\bfg x}_{\sw,{t}}  
 + a{\bfg f_{\ft\ft,y}}\wdt{\bfg y}_{\ft,{t}} +a{\bfg f_{\ft\sw,y}}\wdt{\bfg y}_{\sw,{t}} \\
  &\quad + a^*{\bfg f_{\ft\ft,x}} \wdt{\bfg x}_{\ft,{t+h_\sw}} + a^*{\bfg f_{\ft\sw,x}} \wdt{\bfg x}_{\sw,{t+h_\sw}} 
  + a^*{\bfg f_{\ft\ft,y}} \wdt{\bfg y}_{\ft,{t+h_\sw}} + a^*{\bfg f_{\ft\sw,y}} \wdt{\bfg y}_{\sw,{t+h_\sw}} 
\end{aligned}
\label{eq:p1}
\end{equation}
\begin{equation}
\begin{aligned}
  \wdt{\bfg x}^P_{\sw,{t} + h_\sw} &= \wdt{\bfg x}_{\sw,{t}} + a{\bfg f_{\ft\sw,x}} \wdt{\bfg x}_{\ft,{t}} + a{\bfg f_{\sw\sw,x}} \wdt{\bfg x}_{\sw,{t}}   + a{\bfg f_{\sw\ft,y}}\wdt{\bfg y}_{\ft,{t}} +a{\bfg f_{\sw\sw,y}}\wdt{\bfg y}_{\sw,{t}}   \\
 & \quad + a^* {\bfg f_{\ft\sw,x}} \wdt{\bfg x}_{\ft,{t+h_\sw}} + a^*{\bfg f_{\sw\sw,x}} \wdt{\bfg x}_{\sw,{t+h_\sw}}
 + a^*{\bfg f_{\ft\sw,y}} \wdt{\bfg y}_{\ft,{t+h_\sw}} + a^*{\bfg f_{\sw\sw,y}} \wdt{\bfg y}_{\sw,{t+h_\sw}}
\end{aligned}
\label{eq:p2}
\end{equation}
\begin{equation}
\begin{aligned}
%\hspace{-6.5mm}
\bfg{0}_{m_\ft,1} &= 
\bfg{g}_{\ft\ft,x} \wdt{ \bfg{x}}^P_{\ft,{t} + h_\sw}
+ \bfg{g}_{\ft\sw,x}\wdt{ \bfg{x}}^P_{\sw,t + h_\sw}    + \bfg{g}_{\ft\ft,y} \wdt{\bfg{y}}^P_{\ft,t + h_\sw}  
+  \bfg{g}_{\ft\sw,y} \wdt{ \bfg{y}}^P_{\sw,{t} + h_\sw}  
\end{aligned}
\label{eq:p3}
\end{equation}
\begin{equation}
\begin{aligned}
\bfg{0}_{m_\sw,1} &= \bfg{g}_{\sw\ft,x} \wdt{ \bfg{x}}^P_{\ft,{t} + h_\sw}  
+ {\bfg{g}_{\sw\sw,x}}\wdt{ \bfg{x}}^P_{\sw,{t} + h_\sw}   + {\bfg{g}_{\sw\ft,y}} \wdt{\bfg{y}}^P_{\ft,{t} + h_\sw}  
+  {\bfg{g}_{\sw\sw,y}} \wdt{ \bfg{y}}^P_{\sw,{t} + h_\sw}  
\end{aligned}
\label{eq:p4}
\end{equation}

%Although our previous work demonstrated how to avoid this requirement through sparse matrix reformulation, we retain the assumption herein to achieve more compact and accessible mathematical formulation. 

From \eqref{eq:p3}, we have: 
\begin{equation}
\begin{aligned}
\wdt{\bfg{y}}^P_{\ft, t + h_\sw}  =&  
-\bfg{g}_{\ft\ft,y}^{-1}
( \bfg{g}_{\ft\ft,x}\wdt{\bfg{x}}^P_{\ft,t + h_\sw} + {\bfg{g}_{\ft\sw,x}}\wdt{\bfg{x}}^P_{\sw, t + h_\sw}     
 + \bfg{g}_{\ft\sw,y} \wdt{\bfg{y}}^P_{\sw, t + h_\sw}
)
\label{eq:p5}
\end{aligned}
\end{equation}

%\color{blue1}  
Note that the invertibility of $\mathbf{g}_{\ft\ft,y}$ is used here to facilitate derivation; however, it is not strictly required provided that a sparse matrix formulation is adopted, similarly e.g., to \cite{tzounas2023unified}.
\color{black}
 
Based on \eqref{eq:p1}-\eqref{eq:p5}, we get the expression of $\wdt{\bfg y}^P_{\sw,t+h_\sw}$: 
\begin{align}
\nonumber
   \wdt{\bfg y}^P_{\sw,t+h_\sw} & =   ( -{\mathbf H_3}^{-1}{\mathbf H_1} + \bfb M_1) \wdt{\bfg x}_{\ft,{t}}  + ( - {\mathbf H_3}^{-1}{\mathbf H_1} + \bfb M_2)\wdt{\bfg x}_{\sw,{t}}\\
 &\quad +   \bfb M_3 \wdt{\bfg y}_{\ft,{t}} \! + \! \bfb M_4 \wdt{\bfg y}_{\sw,{t}}  + \bfb M_1^* \wdt{\bfg x}_{\ft,{t+h_\sw}} \! + \! 
\bfb M_2^* \wdt{\bfg x}_{\sw,{t+h_\sw}} \! \label{eq:p6}\\
\nonumber & \quad+ \bfb M_3^*\wdt{\bfg y}_{\ft,{t+h_\sw}} \! + \! \bfb M_4^* \wdt{\bfg y}_{\sw,{t+h_\sw}}
\end{align}
where
%
%\color{blue1}
\begin{align*}
    {\mathbf M_1} \!  & = \! -  {a{\mathbf H_3}^{\!\!-1}{\mathbf H_1}{\bfg f_{\ft\ft,x}} - a{\mathbf H_3}^{\!\!-1}{\mathbf H_2}{\bfg f_{\sw\ft,x}}}  \\
    {\bfb M_2} \! &= \!  - {a{\mathbf H_3}^{\!\!-1}{\mathbf H_1}{\bfg f_{\ft\sw,x}} -a{\mathbf H_3}^{\!\!-1}{\mathbf H_2}{\bfg f_{\sw\sw,x}}} \\
    {\mathbf M_3} \!  & = \!  - {a{\mathbf H_3}^{\!\!-1}{\mathbf H_1}{\bfg f_{\ft\ft,y}} - a{\mathbf H_3}^{\!\!-1}{\mathbf H_2}{\bfg f_{\sw\ft,y}}} \\
    \mathbf M_4 \!  & = \!  - {a{\mathbf H_3}^{\!\!-1}{\mathbf H_1}{\bfg f_{\ft\sw,y}} - a{\mathbf H_3}^{-1}{\mathbf H_2}{\bfg f_{\sw\sw,y}}} 
\end{align*}
 \color{black}
Replace $a$ in $\bfb M_1$ to $\bfb M_4$ as $a^*$ for $\bfb M_1^*$ to $\bfb M_4^*$. 
For $ i = 1 $, combining \eqref{eq:li_2}, \eqref{li_3_b}  and \eqref{eq:p6} establishes the relationship between $ \xys_{{t} + i h_\ft}$, $\xys_{t+(i-1)h_\ft}$ and  $\xys_{{t} + h_\sw}$ over the interval $[t(i-1) h_\ft, t + i h_\ft]$ for the multirate method $(i \leq r-1)$: 
\begin{equation}
    -\Ctdi_i \xys_{ t + h_\sw} + \Ztdi_i \xys_{ t + i h_\ft}  = \Btdi_i \xys_{t + (i-1) h_\ft}
\end{equation}
with 
%
%\color{blue1}
\begin{align*}
\Ctdi_i \!
& =
\frac{i}{r}\begin{bmatrix}
    {\bfg{0}_{n_\ft,n_\ft}} & {\bfg{0}_{n_\ft,n_\sw}} & {\bfg{0}_{n_\ft,m_\ft}} & {\bfg{0}_{n_\ft,m_\sw}} \\
    a^* \bfg{f}_{\sw\ft,x}\! & \! a^* {\bfg{f}_{\sw\sw,x}} &   a^* \!\bfg{f}_{\sw\ft,y} & \!  a^* {\bfg{f}_{\sw\sw,y}} \\
    {\bfg{0}_{m_\ft,n_\ft}} & {\bfg{0}_{m_\ft,n_\sw}} & {\bfg{0}_{m_\ft,m_\ft}} & {\bfg{0}_{m_\ft,m_\sw}} \\
   \bfb{M}_1^* & \bfb{M}_2^* & \bfb{M}_3^* &  \! \bfb{M}_4^*
\end{bmatrix}
\\
\Ztdi_i &=
\bfg I_{\!n_\ft} \!\oplus\! \bfg 0_{n_\sw+m}
\!+\!
(-b \bfg I_{n_\ft}) \!\oplus\!
 \bfg I_{n_\sw} \!\!\oplus\!
 (-b { \bfg I_{m_\ft}}) \!\oplus\!
 \bfg I_{m_\sw}
 \\
& 
%\hspace{1.5cm}
 \times
\begin{bmatrix}
    {\bfg{f}_{\ft\ft,x}} & {\bfg{f}_{\ft\sw,x}} & {\bfg{f}_{\ft\ft,y}} & {\bfg{f}_{\ft\sw,y}} \\
    \bfg 0_{n_\sw,n_\ft} & \bfg I_{n_\sw} & \bfg 0_{n_\sw,m_\ft} & \bfg 0_{n_\sw,m_\sw} \\
    {\bfg{g}_{\ft\ft,x}} & {\bfg{g}_{\ft\sw,x}} & {\bfg{g}_{\ft\ft,y}} & {\bfg{g}_{\ft\sw,y}} \\
    \bfg 0_{m_\sw,n_\ft} & \bfg 0_{m_\sw,n_\sw} & \bfg 0_{m_\sw,m_\ft} & \bfg I_{m_\sw}
\end{bmatrix}
\end{align*}
\begin{align*}
\Btdi_i &=
\bfg I_{\!n_\ft} \!\oplus\! \bfg 0_{n_\sw+m}
\!+  \Ctdi_i \!+ 
(b^* \bfg I_{n_\ft}) \!\oplus\! \bfg I_{n_\sw} \!\!\oplus\!
 (b^*  {\bfg I_{m_\ft}}) \!\oplus\!
 (\frac{i}{r})\bfg I_{m_\sw}
 \\
& 
%\hspace{1.5cm}
\times
\setlength{\arraycolsep}{1pt}
\begin{bmatrix}
    {\bfg{f}_{\ft\ft,x}}  &  {\bfg{f}_{\ft\sw,x}}  & {\bfg{f}_{\ft\ft,y}}  &  {\bfg{f}_{\ft\sw,y}} \\
      a \bfg{f}_{\sw\ft,x}  &  \bfg I_{n_\sw}+ a \bfg{f}_{\sw\sw,x}  &    a \bfg{f}_{\sw\ft,y}  &  a \bfg{f}_{\sw\sw,y} \\
    {\bfg{g}_{\ft\ft,x}}  &  {\bfg{g}_{\ft\sw,x}} &  {\bfg{g}_{\ft\ft,y}}  & {\bfg{g}_{\ft\sw,y}} \\
    -{\mathbf H_3}^{-1}{\mathbf H_1}+ \bfb M_1 &  -{\mathbf H_3}^{-1}{\mathbf H_2}+ \bfb M_2   &    \bfb M_3  &  (\frac{r}{i}-1)\bfg I_{m_\sw} + \bfb M_4
\end{bmatrix}
\end{align*}
 \color{black}
 
At $i=r$, the slow variables are integrated with the \ac{tm}, and the relationship between $ \xys_{{t} + (r-1) h_\ft}$, $\xys_{{t}}$ and  $\xys_{{t} + h_\sw}$ over the interval $[t+(r-1) h_\ft, t + h_\sw]$ for the multirate method: 
\begin{equation}
     \Ztdi_r \xys_{ t + h_\sw}  = \Btdi_r \xys_{ t + (r-1) h_\ft} + \Ctdi_r \xys_{t} 
\end{equation}
with 
%\color{blue1}
%
\begin{align*}
     \Ztdi_r &= 
\bfb E
- 
 (
b
 \bfg I_{{n_\ft}} \! \oplus \! 
 c
 \bfg I_{{n_\sw}} \! \oplus \!
 b
 \bfg I_{{m_\ft}} \! \oplus \!
 c
 \bfg I_{{m_\sw}}
 )
 \bfb A\\
     \Ctdi_r &= 
 \bfg 0_{n_\ft} \oplus \bfg I_{n_\sw} \oplus 
 \bfg 0_{m}
  + 
\bfg I_{n_\ft} \!\oplus\!
c^*\bfg I_{n_\sw} \!\oplus\!
 \bfg I_{m_\ft} \!\oplus\!
 c^* \bfg I_{m_\sw}
 %\\ &
 %\hspace{1.4cm}
 \times
\begin{bmatrix}
 \bfg 0_{n_\ft}   &\bfg 0_{n_\ft,n_\sw}                     &\bfg 0_{n_\ft,m_\ft}  & \bfg 0_{n_\ft,m_\sw} \\
 \bfg{f}_{\sw\ft,x} & \bfg{f}_{\sw\sw,x} & \bfg{f}_{\sw\ft,y} & \bfg{f}_{\sw\sw,y} \\
 \bfg 0_{m_\ft,n_\ft}                     &\bfg 0_{m_\ft,n_\sw}                     &\bfg 0_{m_\ft}  & \bfg 0_{m_\ft,m_\sw} \\
 \bfg{g}_{\sw\ft,x} & \bfg{g}_{\sw\sw,x} & \bfg{g}_{\sw\ft,y} & \bfg{g}_{\sw\sw,y}
 \end{bmatrix}       
     \\
     \Btdi_r &=
\bfg I_{\!n_\ft} \!\oplus\! \bfg 0_{n_\sw+m}
\!+\!
(b^* \bfg I_{n_\ft}) \!\oplus\!
 \bfg I_{n_\sw} \!\!\oplus\!
 (b^*{ \bfg I_{m_\ft}}) \!\oplus\!
 \bfg I_{m_\sw}
% \\
%& 
\times
\begin{bmatrix}
    {\bfg{f}_{\ft\ft,x}} & {\bfg{f}_{\ft\sw,x}} & {\bfg{f}_{\ft\ft,y}} & {\bfg{f}_{\ft\sw,y}} \\
    \bfg 0_{n_\sw,n_\ft} & \bfg 0_{n_\sw} & \bfg 0_{n_\sw,m_\ft} & \bfg 0_{n_\sw,m_\sw} \\
    {\bfg{g}_{\ft\ft,x}} & {\bfg{g}_{\ft\sw,x}} & {\bfg{g}_{\ft\ft,y}} & {\bfg{g}_{\ft\sw,y}} \\
   \bfg 0_{m_\sw,n_\ft} & \bfg 0_{m_\sw, n_\sw} & \bfg 0_{m_\sw,m_\ft} & \bfg 0_{m_\sw}
\end{bmatrix}
\end{align*}
 \color{black}
 
Finally, we arrive to a system in the form of \eqref{eq:dis}, where:
\[
\begin{array}{cc}
\Etdi_r =
\begin{bmatrix}
\Ztdi_r & & \\
\vdots & \ddots & \\
- \Ctdi_1 & & \Ztdi_1
\end{bmatrix} ,
& 
\Atdi_r =
\begin{bmatrix}
\Btdi_r & & \Ctdi_r \\
& \ddots & \\
& & \Btdi_1
\end{bmatrix}
\end{array}
\]

%\section*{References}

%\bibliographystyle{elsarticle-num}
\bibliographystyle{IEEEtran}
\bibliography{references}

\end{document}